\documentclass{aa}  
\usepackage{graphicx}
\usepackage{txfonts}
\usepackage{caption}
\usepackage{textcomp, gensymb}
\usepackage{color}
\newcommand{\mar}[1]{{\color{black}{#1}}}
\newcommand{\petra}[1]{{\color{black}{#1}}}

\newcommand{\xx}[1]{{\color{black}{#1}}}
\newcommand{\xxin}[1]{{\color{black}{#1}}}
\newcommand{\xinn}[1]{{\color{black}{#1}}}
\usepackage{subcaption}
\usepackage{amsmath}
\usepackage{xcolor}
\usepackage[normalem]{ulem}
\usepackage{float}
\usepackage{placeins}
\titlerunning{The Asymmetry of Fornax Cluster Dwarfs}
\authorrunning{X. Xu et al.}
\begin{document}

   \title{Asymmetry at Low Surface Brightness as an Indicator of Environmental Processes in the Fornax Cluster}
   \author{X. Xu\inst{1}, R. F. Peletier\inst{1}, P. Awad\inst{1,2}, M. A. Raj\inst{1}, R. Smith\inst{3}
          }
   \institute{Kapteyn Astronomical Institute, University of Groningen, PO Box 800, 9700 AV Groningen, The Netherlands
         \and
         Bernoulli Institute for Mathematics, Computer Science and Artificial Intelligence, University of Groningen, 9700AK Groningen,
The Netherlands
         \and
             Departamento de Física, Universidad Técnica Federico Santa María, Avenida Vicuña Mackenna 3939, San Joaquín, Santiago de Chile}

   \date{}

  \abstract
   {Dwarf galaxies play an important role in studying the effects of the environment on galaxy formation and evolution. The Fornax cluster having a dense core and strong tidal fields, offers an ideal laboratory to investigate the influence of the cluster environment on the morphology of these galaxies.}
   {In this study, we aim to explore the relationship between the morphology, in particular the asymmetries of galaxies, and their distances to the cluster centre. This is to study the effect of tidal forces and other environmental processes. We do this by investigating the detailed morphologies of a complete magnitude-limited sample of 556 galaxies within the Fornax cluster, spanning a radius range up to 1.75 Mpc, from its central to the outer regions.}
   {For galaxies in the Fornax Deep Survey, we quantified the morphologies of dwarf galaxies using the non-parametric quantities, Asymmetry (A) and Smoothness (S), as part of the CAS system. Unlike previous work, we use \textit{isophotal} CAS-parameters, which are sensitive to the outer parts of galaxies. We constructed the A-r (asymmetry vs. distance to cluster centre) and S-r (smoothness vs. distance to cluster centre) diagrams to investigate the relationship between morphology and distance. Additionally, we examined the effects of asymmetry on magnitude and colour. Furthermore, to better understand the assembly history of the galaxy cluster, we performed a phase-space analysis for Fornax dwarf galaxies, using spectroscopic redshifts and the projected distance from the cluster centre.}
   {We find that dwarf galaxies in the outer regions of the Fornax cluster have higher values of asymmetry compared to other dwarfs in the cluster, indicating a greater degree of morphological disturbances within dwarf galaxies in these regions. We also find that galaxies in the very inner regions are more asymmetric than those further out. The A-magnitude relation reveals a trend where asymmetry increases as galaxies become fainter, and the A-colour relation shows that galaxies with bluer colours tend to exhibit higher asymmetry. \xxin{We do not find any correlations with smoothness, except that smoothness strongly decreases with stellar mass. }\xxin{We propose that the higher asymmetry of dwarfs in the outer regions is most likely caused by ram pressure stripping. As galaxies fall into the cluster, gas is expelled by intracluster winds, causing 'jellyfish-like' tails and leading to star formation not only in the central regions but also along the tails, causing asymmetric features. These asymmetries persist until the galaxies evolve into completely quiescent and elliptical systems. The observed dwarfs likely represent a transitional phase, nearing quiescence but still retaining residual asymmetry from earlier interactions. In the very inner parts, the asymmetries most likely are caused by tidal effects.} In addition, our phase-space diagram suggests that galaxies near pericentre in the Fornax cluster exhibit significantly higher asymmetry, indicating that morphological disturbances happened during their first pericentric passage.}
   {}


   \maketitle
%
%-------------------------------------------------------------------

\section{Introduction}

   The Cold Dark Matter ($\Lambda$CDM) model \citep{1982ApJ...258..415P} is the most successful theory of origin and formation of cosmic structure. According to the $\Lambda$CDM cosmology, larger galaxies grow in a hierarchical way through merger and accretion of smaller systems \citep{1978MNRAS.183..341W,1993MNRAS.264..201K,2006Natur.440.1137S,2007MNRAS.375....2D,2008MNRAS.384....2G}. Dark matter dominates this hierarchical process. During this process, small haloes, the hosts of dwarf galaxies, merge together to finally form the hosts of massive galaxies. In low-mass haloes, dwarf galaxies form their stars slowly \citep[e.g.][]{2024MNRAS.527.9715R}, contrary to their massive counterparts, which form stars much faster. 
   
   \petra{Dwarf galaxies serve as great} laboratories for understanding the formation and evolution of galaxies \citep{1993MNRAS.264..201K}.
   \xinn{The shallow potential well of dwarf galaxies} makes them very sensitive to their surrounding environment which significantly affects their properties. The morphology-density relation \citep[T-$\Sigma$ relation hereafter,][]{1977ApJ...215..401M,1997ApJ...490..577D,1980ApJ...236..351D} is one of the most fundamental correlations between galaxy properties and environment. \citet{1980ApJ...236..351D} first revealed that for bright galaxies, \petra{the fraction of elliptical galaxies increases with the increase of local galaxy density in clusters, in contrast with the fraction of spiral and irregular galaxies which decreases with that trend}. \citet{ 1987AJ.....94..251B} further extended the T-$\Sigma$ relation by showing that the number fraction of dwarf elliptical galaxies increases with the local number density increase, while those of dwarf irregular galaxies decreases. \citet{2021ApJS..252...18T} shows that this T-$\Sigma$ relation is universal.
   
   In a high density environment such as a galaxy cluster, galaxies have interactions with other galaxies, the cluster's gravitational potential, and with the intracluster medium, providing an ideal laboratory to investigate the effects of the environment on galaxies. Various environmental mechanisms have been proposed to explain the morphological transformation and the evolutionary paths that galaxies follow within clusters. These processes include ram-pressure stripping \citep[RPS;][]{1972ApJ...176....1G,1980ApJ...241..928F,1999MNRAS.308..947A}, starvation or strangulation \citep{1980ApJ...237..692L,2008MNRAS.387...79V}, and harassment \citep{1996Natur.379..613M,1998ApJ...495..139M}, major and minor mergers \citep{1977egsp.conf..401T,1992ARA&A..30..705B}, \petra{and} tidal distortion by the cluster potential as galaxies reach the centre \citep{1990ApJ...350...89B,1998ApJ...509..587F}. The relative importance of these mechanisms in the evolution of cluster galaxies remains debated. Mergers and harassment are expected to be more significant in low to intermediate density environments, while ram-pressure stripping and tidal distortion dominate in high-density environments like clusters \citep{2011MNRAS.410.2593M}.

   The morphology of galaxies is one of the essential keys to understanding physics of galaxy formation and evolution. The morphology of galaxies was described by visual impressions for a long time \citep[e.g.][]{1926ApJ....64..321H,1959HDP....53..275D}. With the wide use of photographic plates and Charged Coupled Devices (CCD), quantitative measurements of galaxy morphology became possible \citep[e.g.][]{1981A&A....95..105V,1989ApJ...346L..53F,1996AJ....111.2238P,2009ApJS..182..216K,2014ARA&A..52..291C}. We aim to focus on the morphology of dwarf galaxies. Dwarf galaxies span a wide range of shapes and morphologies. \xinn{For dwarf galaxies, there are a number of classification schemes in use. We use the system of \citet{1994ESOC...49...13B}. In the regime of faint galaxies (\(M_B > -18\), corresponding approximately to a stellar mass of \(3 \times 10^9 M_\odot\)), there are two branches: the high surface brightness and the low surface brightness branch. All objects on the low surface brightness branch are classified as dwarf ellipticals (dE) if they are not star-forming, and as dwarf irregulars (dIrrs) or irregulars (Irr) if they are star-forming. Low-mass objects on the high surface brightness branch, such as M32, are classified as E by \citet{1994ESOC...49...13B} and as compact dwarfs in our classification. This classification system is most commonly used in the current literature \citep[e.g.][]{2006AJ....132.2432L,2019A&A...625A.143V}. \citet{2012ApJS..198....2K} classify all non-star-forming galaxies on the low surface brightness branch as spheroidals (Sph). At the faint end, dwarf galaxies with stellar masses below \(10^5 M_\odot\) are classified as ultra-faint dwarfs \citep[UFDs;][]{2019ARA&A..57..375S}.} \xxin{Early-type dwarfs} are the most common type of galaxy in galaxy clusters \xxin{\citep{1987AJ.....94..251B}} and contain little current star formation. \xxin{Although dEs generally have a simple elliptical morphology, some also contain morphological structures in the inner parts.} \citet{2002A&A...391..823B} found evidence of non-axisymmetric features such as bars, spiral arms, indicative of a disk structure in Virgo dEs. \citet{2006AJ....132.2432L,2007ApJ...660.1186L} performed a statistical analysis on dEs with blue centers in the Virgo Cluster, emphasizing the recent and ongoing star formation in blue central regions. Recently \citet{2021A&A...647A.100S} show that below $M_r$ = -16 mag no early-type dwarf contains components such as bars and bulges, while such components are often found in brighter objects. Star forming dwarfs generally are highly irregular.
   
   There are generally two methods used to quantify morphology of galaxies: parametric methods and the non-parametric methods. The difference between these methods is that non-parametric methods do not assume an analytical function for the galaxy light distribution, and therefore non-parametric morphologies \citep{2004AJ....128..163L,2000ApJ...529..886C,2013MNRAS.434..282F,2016MNRAS.456.3032P} can be applied to any galaxies including irregular galaxies. Parametric methods, on the other hand, fit a galaxy’s light distribution with analytical models such as the widely adopted Sersic model (Sersic 1963), which fits symmetrical Hubble-type galaxies well. The most common non-parametric methods for quantitative morphology used today are $CAS$ (Concentration, Asymmetry and Smoothness or Clumpiness) parameters \citep{1994ApJ...432...75A,1996ApJS..107....1A,2000AJ....119.2645B,2000ApJ...529..886C,2003ApJS..147....1C}. The Gini ($G$) coefficient and $M_{20}$ coefficient are added as a complement to the $CAS$ system \citep{2004AJ....128..163L}. 
   
   Several previous studies have used non-parametric methods, such as $CAS$ parameters, to analyse the morphological properties of dwarf galaxies. \citet{2008MNRAS.385.1374M} performed an analysis of 24 dwarf elliptical galaxies in the Virgo Cluster and in the field based on SDSS g-band images. They quantified structural parameters, including concentration, asymmetry, and clumpiness, and found correlations between age, asymmetry, and clustercentric distance. \citet{2011MNRAS.410.1076P} investigated dwarf galaxies in the Perseus cluster core and those in the outskirts using the $CAS$ parameters. They found that dEs in the cluster outskirts exhibited higher $A$ and $S$ compared to those in the core. 
   
   Galaxy asymmetry was first used by \citet{1995ApJ...451L...1S} as a quantitative way to analyse morphology when characterizing galaxies observed with the \petra{Hubble Space Telescope}. Asymmetry is further extensively used in the literature \citep{1996MNRAS.279L..47A,1996ApJS..107....1A,1996AJ....112..359V,2000AJ....119.2645B}. \citet{2000ApJ...529..886C} found a relation between the asymmetry and galactic radius, which we hereafter refer to as the asymmetry profile. The asymmetry profiles of ellipticals and S0s often have a central peak at very low radii which may be due to the structures such as central disks, dust clouds, and nuclear components at the central regions of galaxies. The asymmetric outer parts can be good tracers of the processes that galaxies undergo, and especially dwarfs, given their low density. The asymmetry of later type galaxies (spirals and irregulars) shows the opposite trend, with asymmetry increasing as a function of radius. In addition to star formation, dust extinction also plays an important role in the morphological disturbances \citep{2007ApJ...659..162T,2008MNRAS.391.1137L,2021ApJ...911..145Y} and therefore affects the measurement of asymmetry parameters in elliptical galaxies. Compared with earlier-type galaxies, the asymmetries of the outer regions of later-type galaxies are higher than the inner regions, partly due to the presence of star-forming disks.

   The Fornax cluster is the second most massive concentration of galaxies after the Virgo cluster within 20 Mpc, with an estimated mass of $7\times10^{13} M_{\odot}$ \citep{2001ApJ...548L.139D}. The Fornax cluster, due to its relatively close proximity and composition of both a main cluster and an in-falling subgroup, is an ideal laboratory to investigate the formation and evolution of galaxies in cluster environment, which has been well studied by deep multi-wavelength surveys: The Fornax Cluster catalogue \citep[FCC;][]{1989AJ.....98..367F}, the \mar{ACS Fornax Cluster Survey \citep[ACSFCS;][]{2007ApJS..169..213J}, the Next Generation Fornax Survey \citep[NGFS;][]{2015ApJ...813L..15M,2018ApJ...855..142E}}, the Fornax Deep Survey \citep[FDS;]{2016ApJ...820...42I,2018A&A...620A.165V,2020arXiv200812633P}, the ALMA Fornax Cluster Survey \citep{2019MNRAS.483.2251Z} and the MeerKAT Fornax Survey \citep{2023A&A...673A.146S}. In this paper we will consider data from the FDS, an ultradeep survey of 26 square degrees, in 4 bands, u,g,r and i. Several papers have used this survey to study dwarf galaxies \citep[e.g.][]{2019A&A...625A.143V,2022A&A...662A..43V,2021A&A...647A.100S,2020A&A...639A.136C}, and also the outer regions of massive galaxies \citep[e.g.][]{2019A&A...628A...4R,2016ApJ...820...42I,2019A&A...623A...1I,2020A&A...639A..14S}. The upcoming surveys like Euclid \citep{2011arXiv1110.3193L} will be able to provide new observations with higher resolution of the Fornax cluster, which are deeper and of higher spatial resolution than ground-based telescopes.
   
   Previous studies focusing on morphology in the Fornax cluster have revealed the strong influence of the surrounding environment on the morphology of dwarf galaxies. \citet{2019A&A...625A.143V} observe an increasing early-type fraction and a decline in late-type galaxies towards the inner regions of the cluster. \citet{2021A&A...647A.100S} found significant differences in various structural properties, including colours, structural parameters, and non-parametric morphological parameters (Concentration C; Asymmetry A, Clumpiness S; Gini; M20), between galaxies in the Fornax main cluster and Fornax A group. Fornax A galaxies have bluer colours, smaller sizes, and higher A and S compared to galaxies in the Fornax main cluster. Moreover, they found that in the Fornax main cluster, galaxies becoming fainter and more extended towards the cluster centre.
   
   In this paper, we use data from the FDS to investigate the morphological characteristics of dwarf galaxies within the Fornax cluster, spanning its central to outer regions. Our study focuses on the effects of the local environment on the morphology of dwarf galaxies in galaxy clusters. We use the asymmetry and smoothness parameters as methods to quantify the morphologies of these dwarf galaxies. Specifically, we explore the relationship between the morphology of the dwarfs and their distances to the cluster centre. This paper is structured as follows: In Section~\ref{sec:obsData}, we provide a description of our observational data and catalogue. We then introduce the non-parametric CAS system used for quantifying the morphology of our dwarf galaxies in Section~\ref{sec:methods}. In Section~\ref{sec:results}, we present \petra{our results in terms of the} morphology-position relation, the A-magnitude relation and the A-colour relation. \petra{Our} results, including the phase-space diagram analysis, are \petra{then} discussed in Section~\ref{sec:discussion}. Finally, we \petra{conclude and} summarize our results in Section~\ref{sec:conclusion}.
%--------------------------------------------------------------------
\section{Observational Data}
\label{sec:obsData}
   The Fornax Deep Survey (FDS) is a joint effort combining NOVA's Guaranteed Time Observations from the Fornax Cluster Ultra-deep Survey (FOCUS, P.I. R. Peletier) and the VST Early-type GAlaxy Survey (VEGAS, P.I. E. Iodice). This deep, multi-band imaging survey covers a large area of 21 deg$^2$ in the $u$, $g$, $r$, and $i$ bands centered on the whole Fornax galaxy cluster, and an additional 5 deg$^2$ in the $g$, $r$, and $i$ bands centered on the Fornax A group. The data details and reduction procedures are described in papers by \citet{2018A&A...620A.165V}. The Fornax Deep Survey is one of the new imaging surveys which provides a deep, multi-band ($u’,g’,r’,i’$) data set to study the faintest dwarf galaxies within the Fornax cluster. The data was collected with OmegaCAM \citep{2002Msngr.110...15K} at the ESO VLT Survey Telescope (VST) \citep{2012SPIE.8444E..1CS}.
   
   We used the Fornax Deep Survey Dwarf Galaxy catalogue (FDSDC) published by \citet{2018A&A...620A.165V}, which includes 564 dwarf galaxies, that were detected with SExtractor \citep{1996A&AS..117..393B}. The catalogue defines dwarfs as galaxies with an r-band absolute magnitude fainter than $\mathrm{M}_{r^{\prime}}=-18.5~\mathrm{mag}$. It reaches a 50\% completeness limit at $\mathrm{M}_{r^{\prime}}=-10.5~\mathrm{mag}$ and a limiting mean effective surface brightness of $\bar{\mu}_{e, r^{\prime}}=26~\mathrm{mag} \operatorname{arcsec}^{-2}$. We excluded eight duplicates from this catalogue subsequently. 
   Information about the galaxies in the FDSDC is also given in the photometric catalogue compiled by \citet{2021A&A...647A.100S}, who investigated the structural properties of galaxies in the Fornax main cluster and the infalling Fornax A group using data from the FDS. \citet{2021A&A...647A.100S} quantify the light distribution of each galaxy using a combination of aperture photometry, Sérsic+point spread function, and multi-component decompositions, as well as non-parametric measures of morphology. This catalogue contains a total of 582 galaxies and provides essential information such as position angle, effective radius, and axial ratio for each galaxy.

   %We did not include the additional 265 galaxies presented in \citet{2022A&A...662A..43V}. Therefore, our final sample consisted of 556 dwarf galaxies. 
%--------------------------------------------------------------------
\section{Methods}
\label{sec:methods}
%-------------------------------------- Two column figure (place early!)
   \begin{figure*}
   \centering
   \begin{subfigure}[b]{0.45\textwidth}
      \centering
      \includegraphics[width=0.78\textwidth]{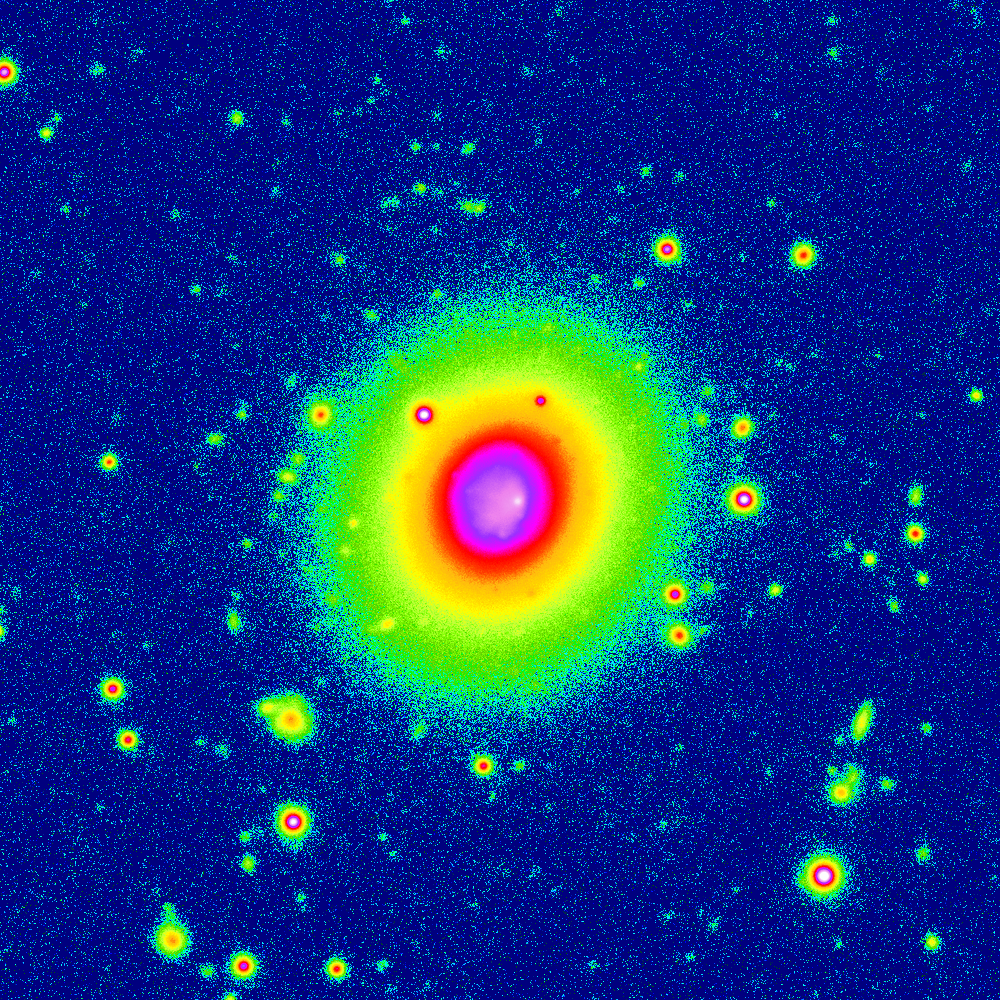}
   \end{subfigure}
   \hfill
   \begin{subfigure}[b]{0.5\textwidth}
      \centering
      \includegraphics[width=\textwidth]{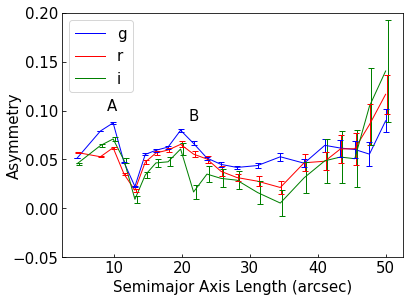}
   \end{subfigure}
   \begin{subfigure}[b]{0.45\textwidth}
      \centering
      \includegraphics[width=0.78\textwidth]{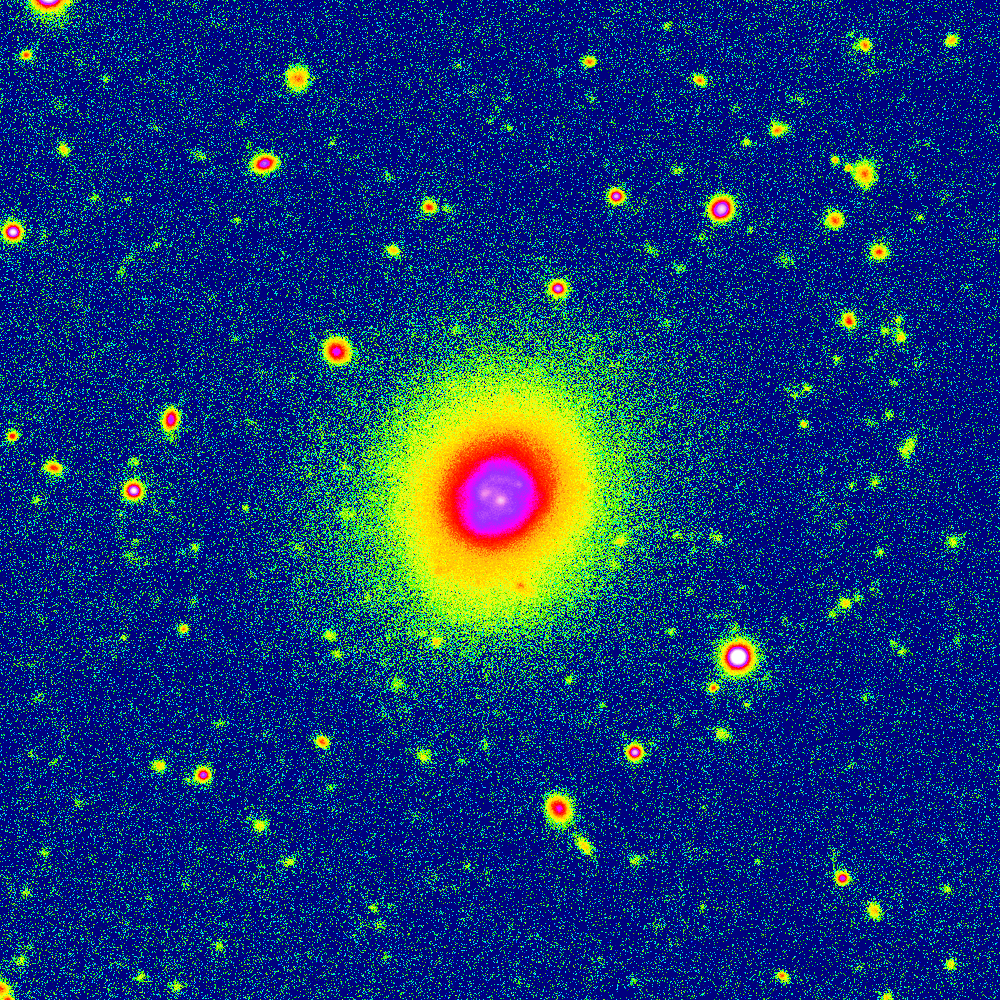}
   \end{subfigure}
   \hfill
   \begin{subfigure}[b]{0.5\textwidth}
      \centering
      \includegraphics[width=\textwidth]{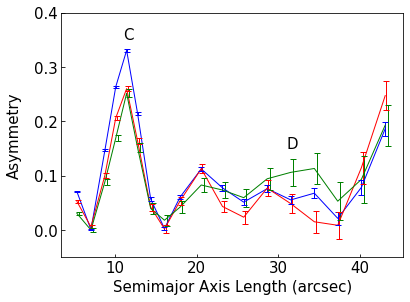}
   \end{subfigure}
   \caption{Two representative dwarf galaxies from the FDS sample. upper panel: r-band image of dwarf galaxy F4D000 (left) and the corresponding asymmetry parameter as a function of radius in the g, r, and z bands (right). \xxin{The error bars represent the asymmetry uncertainties.} bottom panel: r-band image of dwarf galaxy F4D053 alongside the asymmetry profile.}
   \label{fig:Aprof}
   \end{figure*}

   %\subsection{Morphology}  
   %\subsection{Nonparametric morphology indicators \sout{(Morphological measurements)}}   
   In this study, we quantify the morphologies of our sources using two \petra{of these} non-parametric measures, \petra{namely} asymmetry and smoothness. We decided not to use the concentration throughout this study, as it does not effectively show irregularities in a galaxy’s light distribution, which are the crucial aspects of our analysis. Below we provide a brief description of the basic details for each of the parameters. 

   \begin{description}
   \item[\textbf{Asymmetry (A)}:]
   The asymmetry parameter quantifies the degree of rotational symmetry of light in a galaxy. The asymmetry of a galaxy is obtained by rotating a galaxy by 180\degree and then subtracting the rotated image from the original image. Asymmetry is calculated as \citep{1995ApJ...451L...1S,2000ApJ...529..886C,2003ApJS..147....1C}:\\
   \begin{equation}
   A=\frac{\sum\left(I_{0}-I_{\theta}\right)}{2 \sum\left|I_{0}\right|}-A_{b g r}
   \label{equ:asy}
   \end{equation}
   Where $I_0$ is the original galaxy image, $I_\theta$ is the galaxy image rotated by 180 degrees about a chosen centroid which is determined by minimizing $A$, and $A_{bgr}$ is the average asymmetry in the background. The sum is over all pixels within $1.5 \times r_{petro}$ of the galaxy’s centre. The center of rotation is decided by an iterative process which makes an initial guess of the center to compute \petra{A} and then alters the center until the location of the minimum asymmetry \petra{is found} \citep{2000ApJ...529..886C}. The asymmetry can be used to identify ongoing galaxy interactions and mergers \citep{2000ApJ...529..886C,2003ApJS..147....1C}, and its \petra{correlation} with the colour of a galaxy has been shown in previous studies \citep{1995ApJ...451L...1S,2000A&A...354L..21C,2000ApJ...529..886C,2003ApJS..147....1C}\\

   \item[\textbf{Smoothness/Clumpiness (S)}:]
   Smoothness or Clumpiness measures the degree of small-scale structure and describes the patchiness of the light distribution in a galaxy \citep{2003ApJS..147....1C}. \petra{S} is defined by \citet{2003ApJS..147....1C} as the ratio of the amount of light at high spatial frequencies to the total light of a galaxy. \petra{This value} is obtained by smoothing the original galaxy image with a boxcar filter of a given width and then subtracting the blurred image from the original image. It can be calculated as:\\
   \begin{equation}
   S=10 \times \sum_{x, y=1,1}^{N, N} \frac{\left(I_{x, y}-I_{x, y}^{\sigma}\right)-B_{x, y}}{I_{x, y}}
   \label{equ:smooth}
   \end{equation}
   where $I_{x,y}$ is the flux values of the galaxy at pixel \petra{with position} ($x,y$),  \petra{$I_{x,y}^{\sigma}$} is the pixel flux values of the image smoothed using a boxcar of width $0.25\times r_{petro}$, and $B_{x,y}$ is the pixel values of a smoothed background region. The sum is carried using all pixels within $1.5\times r_{petro}$ but pixels with $0.25 \times r_{petro}$ are excluded to avoid the highly concentrated central regions of galaxies. \citet{2003ApJS..147....1C} showed that \petra{measuring the} clumpiness \petra{of a galaxy} can be a good method for \petra{studying} star-formation, \xxin{as newly formed stars are found in clumpy star-forming regions, making the clumpiness of a galaxy’s light a natural tracer of young stars}. Higher value of $S$ means \petra{that} there are clumps within a galaxy, hence elliptical galaxies generally should have \petra{a negligible} $S$. Spiral galaxies and irregular galaxies, \petra{on the other hand,} contain many clumps due to the presence of star forming regions \petra{and therefore have a high S estimate}.

   \end{description}
   
    \begin{figure*}
   \centering
   \includegraphics[width=2\columnwidth]{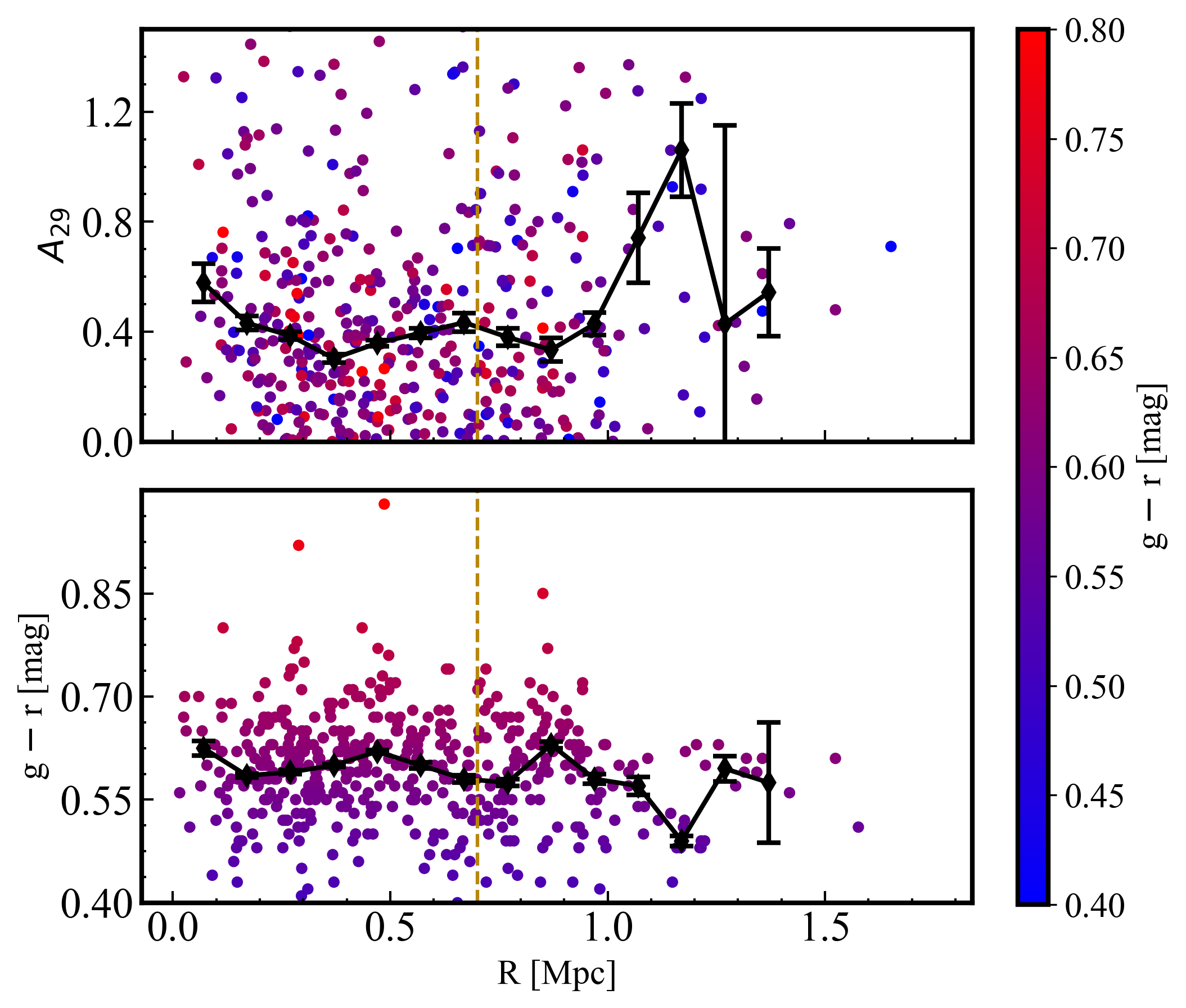}
      \caption{The dependence of parameter $A$ on the projected cluster-centric radius $R$ (upper panel) \xx{and the variation of colour with $R$ (lower panel)}. The black dots represent the median values of the measurements in each bin, \xxin{with the error bars indicating the bootstrapped 95\% confidence interval.} The colour bar corresponds to the $g-r$ colour.%The right panel shows the corresponding number fractions as function of radius. Dwarf galaxies with higher asymmetry and those with lower asymmetry are represented in different colours, The corresponding legend can be found in the right panel. The blue solid line at 0.36 Mpc indicates a high-density region, and 
       The yellow dashed line at 0.7 Mpc indicates the virial radius.
      }
         \label{fig:A_r}
   \end{figure*}

   \subsection{Asymmetry profile}   
   In order to investigate the outer regions of galaxies, specifically to detect possible indications of interactions between galaxies or tidal stripping, we analyse the asymmetry profiles of all available Fornax dwarfs. We determine the central position of each galaxy by calculating their centroid using the \texttt{photutils.morphology} package \citep{2021zndo...5525286B}. The process involved the following steps: a segmentation map was generated using the \texttt{detect\_sources} function. To identify the main galaxy segment within each image, the central coordinates of the image were used. Relevant data properties were then extracted from the selected galaxy segment and image. Using the \texttt{cat.centroid} function from the extracted catalogue, we calculated the precise central position of each galaxy. The centroid is computed as the center of mass of the unmasked pixels within the source segment. We used the initial centroid to build an isophote at 27 mag/arcsec$^2$, and the center of this isophote was determined as the final centroid.
   
   Additionally, we use the python \texttt{astropy.photutils}\footnote{https://photutils.readthedocs.io/en/stable/index.html} package to fit elliptical isophotes on our images with the position angle and ellipticity fixed. The asymmetry is computed in each elliptical annulus. Fig.~\ref{fig:Aprof} presents two representative examples of asymmetry profiles in \petra{the $g$, $r$, and $i$} passbands of two galaxies within the Fornax cluster. The upper panels represent dwarf F4D000 , while the lower panels represent dwarf F4D053. In the upper panel, the asymmetry in the $g$-band is higher than in the $r$ and $i$ bands within regions A and B, attributed to the influence of dust extinction. In the lower panel, we again observe higher $g$ band asymmetry compared to $r$ and $i$ in region C, due to dust extinction. In region D, the asymmetry in the $i$ band exceeds that in the $g$ and $r$ bands, potentially because of more prominent fringes in the $i$ band images. The asymmetry profiles vividly show the morphological features of the two galaxies, enabling us to understand how the surrounding environment influences the morphology of dwarf galaxies in the cluster.
   
   Given that \citet{2021A&A...647A.100S} study the general asymmetry as a function of projected cluster-distance and colour but did not observe a strong trend, we instead focus on asymmetry in the outskirts. To quantify the asymmetry in the outskirts of a galaxy, we focus on asymmetry within the magnitude range of 26-29 mag/arcsec$^2$, which we denote by: A26 - between isophotes of 25.5 and 26.5, A27 - between isophotes of 26.5 and 27.5, A28 - between isophotes of 27.5 and 28.5, and A29 - between isophotes of 28.5 and 29.5. This range was chosen with consideration for the high susceptibility of the outskirts of galaxies to their surrounding environment. Our investigation aims to show how environmental factors influence the asymmetry of these outer regions. In addition to calculating the asymmetry, we determined the smoothness by employing the same method.
   
   \subsection{Background Subtraction}
   Even though the sky background had been subtracted, we still need to get rid of small gradients from the image. Therefore, we proceeded with the following major steps: 
   \begin{itemize}
       \item Source masking: all galaxies and stars were detected and masked in the image.
       \item Sky background model fitting: a functional model $z = C_1(x-x_0) + C_2(y-y_0) + C_0$ was used to fit the sky background on each masked image using a least-squares fit, where $z$ represents the pixel intensity, $x$ and $y$ are the pixel coordinates within the image, $x_0$ and $y_0$ are the center coordinates of the image, and the coefficients $C_0$, $C_1$, and $C_2$ were determined through the fitting procedure.
       \item Sky background subtraction: the fitted sky background model was subtracted from the original image, which eliminates the contribution of the sky background.
   \end{itemize}

%--------------------------------------------------------------------
\section{Results}
\label{sec:results}

   In this section, we determine the asymmetry and smoothness of 556 dwarf galaxies within the Fornax Cluster. Before computing the CAS parameters, we remove background galaxies and foreground stars using a mask generated by \citet{2021A&A...647A.100S}. Late-type galaxies show inherent irregularities that can complicate the differentiation between asymmetries induced by environmental factors and those arising from their natural characteristics. To investigate the effect of environmental processes on galaxies, in the following analysis, we only concentrate on dwarf galaxies classified as early-type galaxies.
   
   The properties of cluster galaxies, such as their morphologies, are closely linked to the local galaxy density (\ensuremath{T - \Sigma \text{ relation}}), 
   which is strongly decreasing as a function of the distance from the cluster centre ($T-R$ relation) \citep{1993ApJ...407..489W,2015MNRAS.449.3927F,2020A&A...640A..30M}. 
   
   In order to examine the relationship between the asymmetries of our dwarf galaxies and their distances from the cluster centre, we calculate the projected distance of each dwarf galaxy from the cluster centre and plotted asymmetries and smoothness as a function of galactocentric distance. Here we use NGC 1399 as the cluster centre \citep{2018A&A...620A.165V}.

   To compute the uncertainties associated with the asymmetry measurements, we used a subsample of 25 galaxies randomly chosen from the FDS to investigate the flatness of the sky and the accompanying errors, which provided a minimum uncertainty value of 0.002. This value was then used to estimate the photometric error, the primary source of uncertainty in our asymmetry measurements. \xinn{The uncertainties in the asymmetry of each dwarf galaxy were calculated using the photometric error and Poisson statistics.} These asymmetry uncertainties were then used to calculate the error-weighted asymmetries across the four regions of the cluster, as presented in Table 1. To quantitatively assess the relation between asymmetry (A) and the cluster environment, we calculated the median and the 95\% confidence interval in these four distinct regions: the very central region  (\(R \leq 0.12 \, \text{Mpc}\)), a central region (\(0.12 < R \leq 0.4 \, \text{Mpc}\)), an intermediate region (\(0.4 < R \leq 1 \, \text{Mpc}\)) and an outer region (\(R > 1 \, \text{Mpc}\)).
   \subsection{The radial distribution of asymmetry}
   Figure~\ref{fig:A_r} shows how asymmetry and colour vary with the projected distance from the Fornax cluster centre. The colour bar represents the \(g-r\) colour in the $r$ band for each galaxy. The data are divided into equal bins (0.1 Mpc) based on the distance from the cluster centre. \xxin{We used the bootstrapping method with 9999 resamples for each radial bin, and the error bars represent the 95\% confidence intervals based on the scatter of the asymmetry values within each bin. }As can be seen,  dwarf galaxies in the outer region ($R \gtrsim 1 \text{ Mpc}$) of the Fornax cluster exhibit  higher values of $A$ compared to other dwarfs. \xxin{As shown in Table 1, the asymmetry is systematically higher in the outer region ($R > 1 \text{ Mpc}$) compared to the intermediate region ($0.4 < R \leq 1 \text{ Mpc}$). In the very central region ($R \leq 0.12 \text{ Mpc}$), asymmetry is similarly higher  than in the region just outside it ($0.12 < R \leq 0.4 \text{ Mpc}$), indicating that both the outer and very central regions have statistically significant higher asymmetry compared to the intermediate region.} 
   
   Additionally, we find that the \(g-r\) colour slowly bluens towards the outer regions, suggesting that star formation is driving the enhanced asymmetry observed there. \xxin{In contrast, in the inner regions, the \(g-r\) colour shows a radial gradient towards becoming more red, indicating that the increasing asymmetry cannot be attributed to an increasing number of star-forming regions and may instead be influenced by tidal forces affecting this area.}
    \begin{table*}
    \centering
    \resizebox{\textwidth}{!}{
    \begin{tabular}{c|cc|cc|cc|cc}
        \hline
         & \multicolumn{2}{c|}{Very Central} & \multicolumn{2}{c|}{Central} & \multicolumn{2}{c|}{Intermediate} & \multicolumn{2}{c}{Outer}
        \\ \hline
        Asymmetry & \(R \leq 0.12 \, \text{Mpc}\) & uncertainty & \(0.12 < R \leq 0.4 \, \text{Mpc}\) & uncertainty & \(0.4 < R \leq 1 \, \text{Mpc}\) & uncertainty & \(R > 1 \, \text{Mpc}\) & uncertainty\\ \hline
A26 & 0.0995&	0.0091&	0.0559&	0.0006&	0.0695&	0.0005&	0.0829&	0.0033 \\  
A27 & 0.1882&	0.0222&	0.0966&	0.0010&	0.1037&	0.0007&	0.1052&	0.0044 \\  
A28 & 0.4918&	0.0492&	0.1953&	0.0022&	0.2079&	0.0014&	0.3399&	0.0165 \\  
A29 & 0.6295&	0.0420&	0.4496&	0.0046&	0.4213&	0.0029&	0.7156&	0.0331 \\ \hline
    \end{tabular}}
    \end{table*} 
    \begin{table*}
    \centering    
    \begin{tabular}{c|cccc}
        \hline
         & Out-Int & Very Cen-Cen\\ \hline
A26 & 0.0134	$\pm$	0.0101&	0.0436	$\pm$	0.0275 \\  
A27 & 0.0015	$\pm$	0.0134&	0.0916	$\pm$	0.0665 \\  
A28 & 0.1320	$\pm$	0.0495&	0.2965	$\pm$	0.1478 \\  
A29 & 0.2943	$\pm$	0.0996&	0.1799	$\pm$	0.1269 \\ \hline 
    \end{tabular}
    \caption{\xxin{Upper Panel: Weighted Mean and weighted standard error of the mean for asymmetry (A) in the Very central, Central, Intermediate, and Outer Regions of the Fornax Cluster (\(R \leq 0.12 \, \text{Mpc}\), \(0.12 < R \leq 0.4 \, \text{Mpc}\), \(0.4 < R \leq 1 \, \text{Mpc}\), \(R > 1 \, \text{Mpc}\)). Lower Panel: Differences between the weighted means for the regions (Out-Int and Very Cen-Cen), along with the half-widths of the 95\% confidence intervals.}}
    \label{tab:asymmetry}
    \end{table*} 
    
\begin{figure}
   \centering
   \includegraphics[width=1\columnwidth]{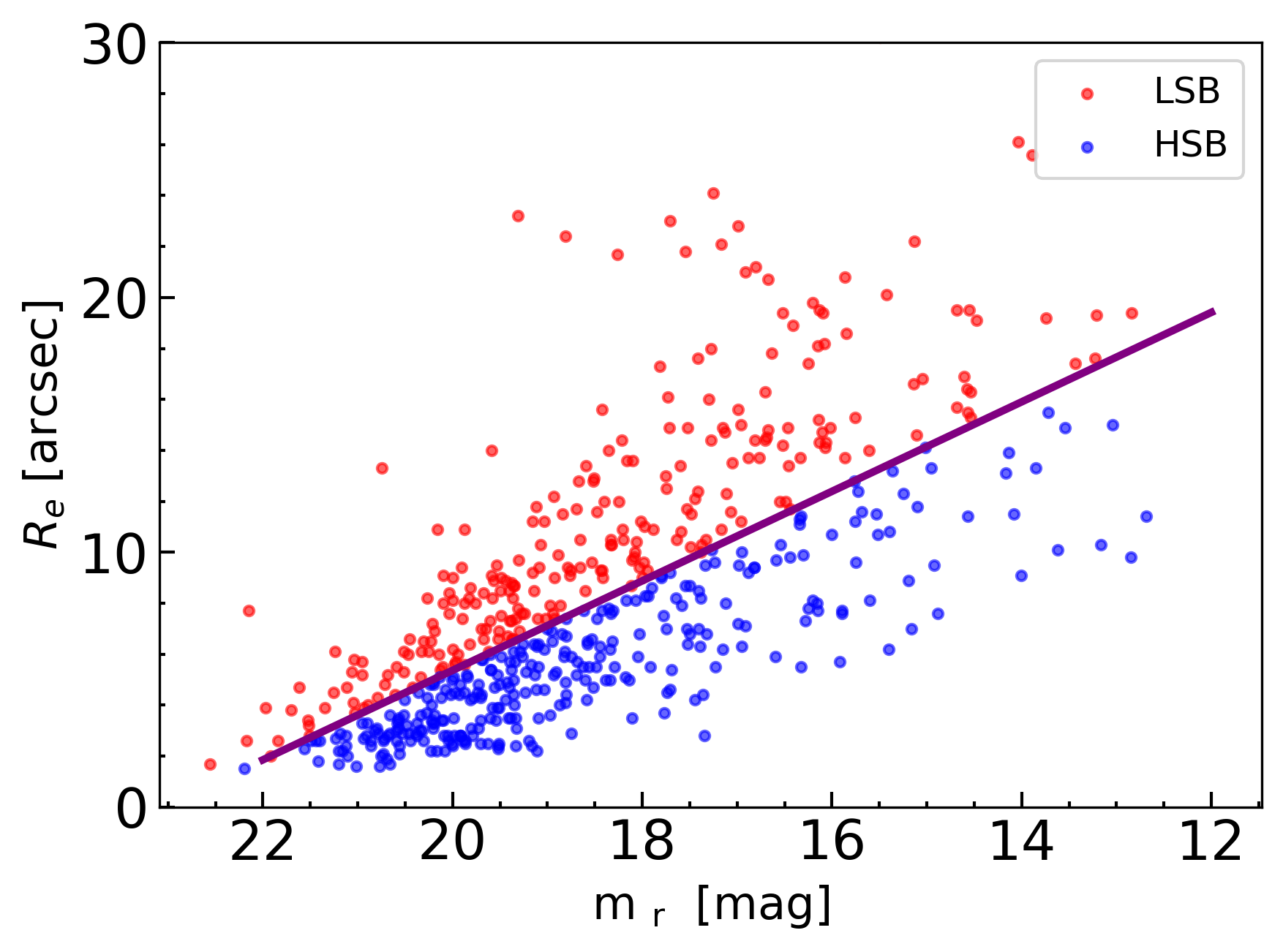}
      \caption{\xxin{Effective radius (\(R_e\)) versus \(r\)-band magnitude for Fornax dwarf galaxies. We define the galaxies below the best-fitting relation as HSB galaxies (blue) and the ones above as LSB galaxies (red).}}
         \label{fig:Re_r}
   \end{figure}

\begin{figure}
   \centering
   \includegraphics[width=1\columnwidth]{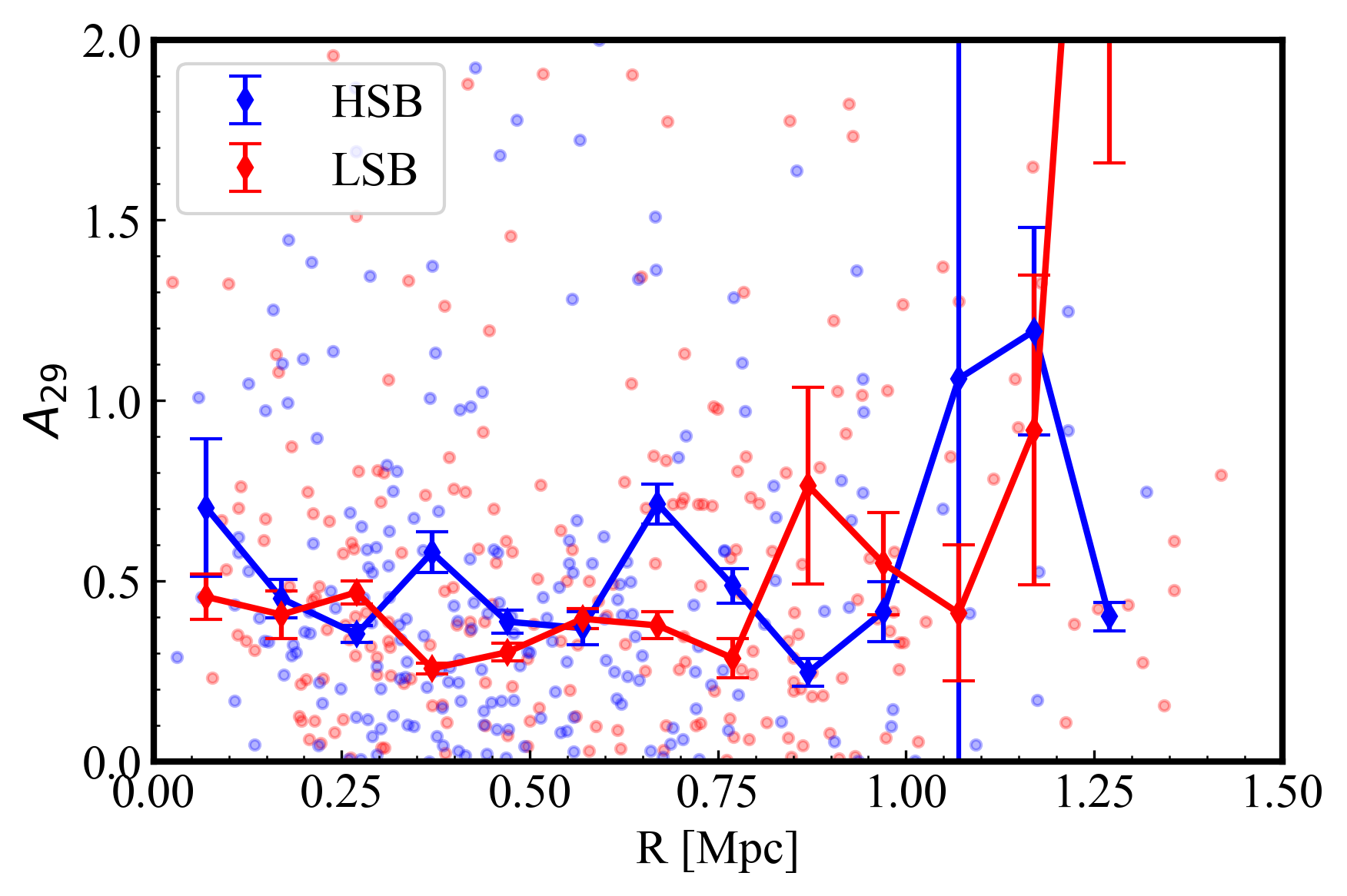}
      \caption{\xxin{Same as Fig.~\ref{fig:A_r} for the \(A\)–\(R\) relation. HSB galaxies are shown in blue, and LSB galaxies are shown in red.}}
         \label{fig:A_lsbhsb}
   \end{figure}

   Low surface brightness galaxies are more susceptible to environmental effects, such as harassment and tidal stripping, compared to high surface brightness galaxies. To investigate this phenomenon, \xxin{we made a test, and split the sample in a lower surface brightness (LSB) and a higher surface brightness (HSB) sample.} \xxin{Figure~\ref{fig:Re_r} shows the effective radius as a function of \(r\)-band magnitude for all dwarf galaxies. Dwarfs located above the best-fitting line (\(R_e = -1.8r + 41.466\)) are classified as LSB galaxies, while those below the line are classified as HSB galaxies.}
   We examined whether asymmetry correlates with surface brightness\xxin{, as shown in Fig.~\ref{fig:A_lsbhsb},} and found no clear correlation. \xxin{This suggests that the difference in surface brightness between the two samples for a given luminosity is not very large, indicating that asymmetry may be more dependent on other factors, such as galaxy size, rather than surface brightness.}
   
\begin{figure*}
   \centering
   \includegraphics[width=2\columnwidth]{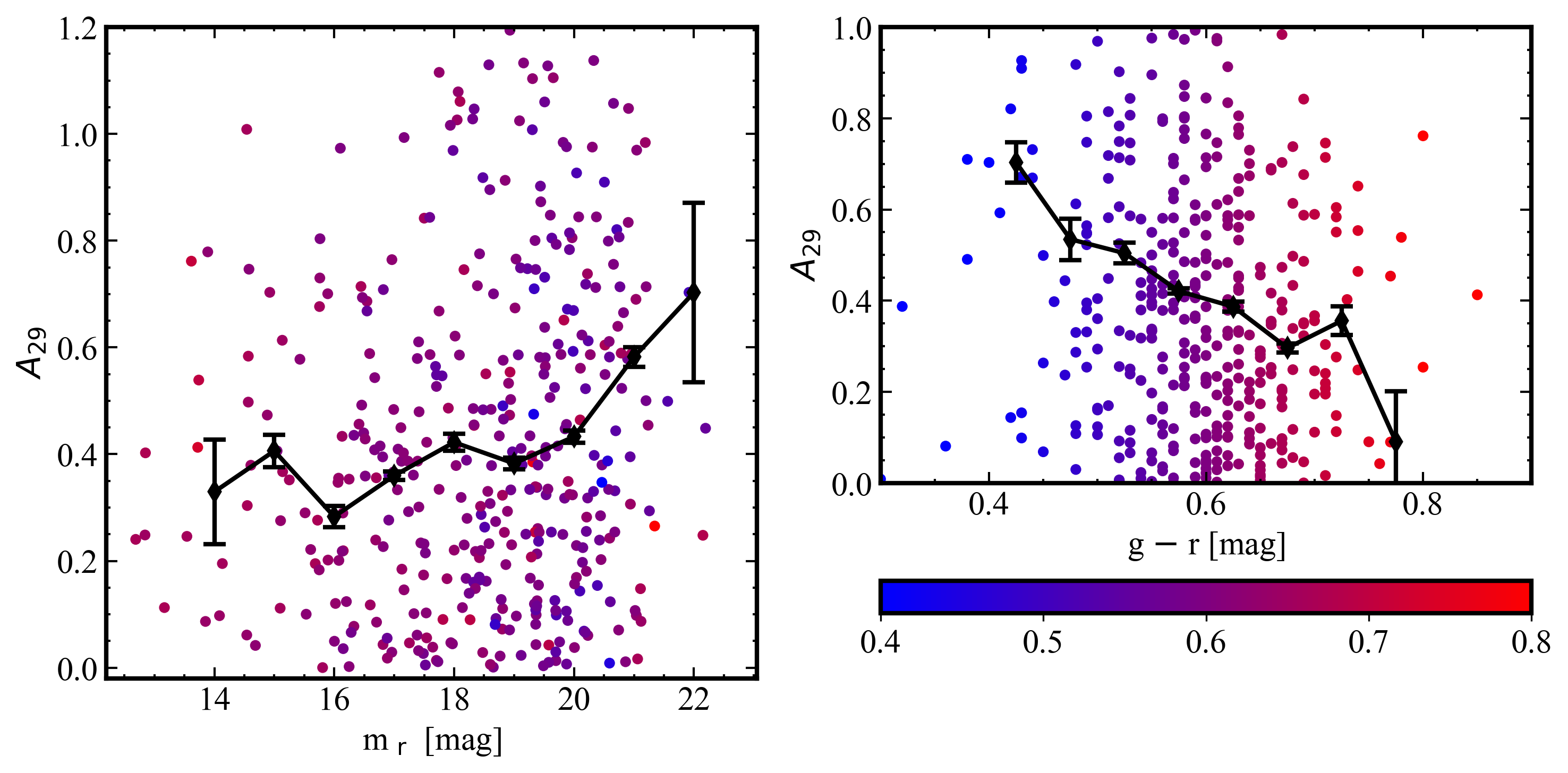}
      \caption{Left panel: Asymmetry-magnitude relation for dwarf galaxies in the Fornax cluster. The colour bar represents the g-r colour. Right panel: Relationship between asymmetry and $g-r$ colour. The error bar represents the variation in colour.}
         \label{fig:A_mag_colour}
   \end{figure*}

   Furthermore, we also explore the relationship between asymmetry and magnitude in the left panel of Fig.~\ref{fig:A_mag_colour}. The figure reveals an increasing trend as the magnitude becomes fainter. This trend suggests that faint galaxies are more susceptible to the effects of their surrounding environment, resulting in higher asymmetry values. The trend may also be influenced by the fact that fainter, generally bluer galaxies are often younger and have more clumps, contributing to their increased asymmetry. Additionally, we observe slightly higher asymmetry in bright galaxies, which suggests that it may be caused by dust. We further present the relationships between asymmetry and distance, and between asymmetry and magnitude, for additional asymmetry measures (A26-A29) in Appendices A and B, respectively.
   \subsection{A vs colour}
   The right panel of Fig.~\ref{fig:A_mag_colour} presents the correlation between asymmetry and colour, focusing on dwarf elliptical galaxies. The results indicate that bluer \xxin{dEs} tend to be more asymmetric. This asymmetry-colour relationship is consistent with the findings of previous authors \citep[e.g.][]{1997PASP..109.1251C,1999Ap&SS.269..585C,2007ApJ...659..162T,2000AJ....119.2645B}. Asymmetric galaxies tend to have bluer colour, indicating the presence of a small amount of recently formed massive young stars and dust, given that the object is still classified as dE. In contrast, symmetric galaxies have a redder colour, suggesting a relatively older stellar population.

%--------------------------------------------------------
\subsection{S vs magnitude}
\begin{figure}
   \centering
   \includegraphics[width=1\columnwidth]{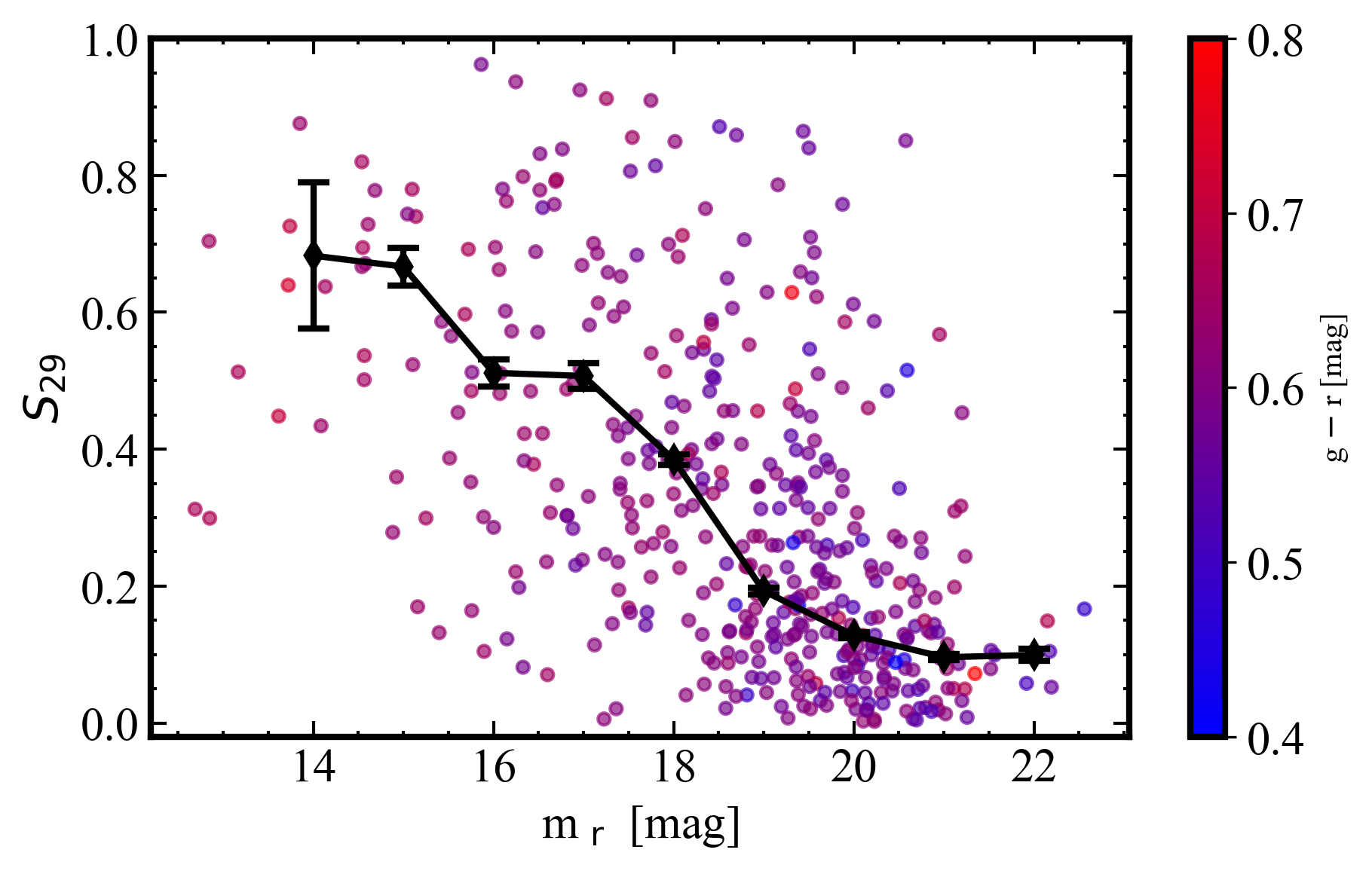}
      \caption{Same as Fig.~\ref{fig:A_mag_colour} for the S – magnitude relation. }
         \label{fig:S_mag}
   \end{figure}

Figure~\ref{fig:S_mag} shows a decreasing trend in S29 as the magnitude becomes fainter. This trend can be attributed to the fact that faint \xxin{dEs} generally have featureless outskirts. The lack of substructure in these regions results in a smoother appearance, hence the observed decrease in smoothness with increasing magnitude. \xxin{Similarly, \citet{2021A&A...647A.100S} find that smoothness decreases with decreasing stellar mass.} \xx{Furthermore, research by \cite{2022AJ....164...18M} indicates that brighter dwarf ETGs tend to have a higher probability of possessing embedded substructures, such as disk or clump features. Therefore, the trend in smoothness may indeed reflect these intrinsic differences.}

\section{Discussion}
\label{sec:discussion}

 Our study reveals that dwarf galaxies in the outer region have higher values of asymmetry (A) compared to dwarfs \petra{in other areas of the cluster}. Additionally, while we observe a subtle asymmetry among dwarf galaxies in the central region of the Fornax cluster, there is no clear trend indicating consistently higher asymmetry values compared to dwarfs in other areas of the cluster. Following these results, we will discuss three aspects in the following subsections: the effect of the surrounding environment on dwarf galaxies in the galaxy cluster (Sect.5.1), the application of the projected phase-space (PPS) diagram for understanding galaxy evolution (Sect.5.2), and an analysis of the inner region related to the observed subtle asymmetry in galaxies within this densest area of the Fornax cluster (Sect.5.3).
   
   \subsection{The effect of local environment on galaxy morphology}
   For galaxies in the outskirts of the Fornax cluster, our results reveal that they are more morphologically disturbed on average, which is consistent with previous studies. \citet{2011MNRAS.410.1076P} quantified the morphologies of dwarf galaxies in the Perseus Cluster using the CAS systems and found that apparent \xxin{dEs} in the outskirts of Perseus have higher values of both A and S compared to those in the cluster centre. These results indicate that the morphology of dwarf galaxies becomes increasingly disturbed as their distance from the cluster centre increases. 
   
   The disturbed morphologies of galaxies everywhere in the cluster can be explained by physical processes related to the surrounding environment, such as ram pressure stripping and harassment. The strength of ram pressure stripping scales with $\rho_{\text{ICM}} v_{\text{rel}}^2$, where $\rho_{\text{ICM}}$ is the density of the intracluster medium (ICM) and $v_{\text{rel}}$ is the relative velocity between the satellite galaxy and the ICM \citep{1972ApJ...176....1G}. As a result, galaxies at smaller distances from the cluster centre experience stronger ram pressure stripping due to the higher density of the ICM. This leads to more rapid removal of gas and earlier quenching of star formation \citep{2006MNRAS.369.1021M,2022A&ARv..30....3B}. In contrast, galaxies in the outer regions are more morphologically disturbed as a result of the more recent cessation of star formation.  
   \begin{figure*}[htbp]
   \centering
   \includegraphics[width=2\columnwidth]{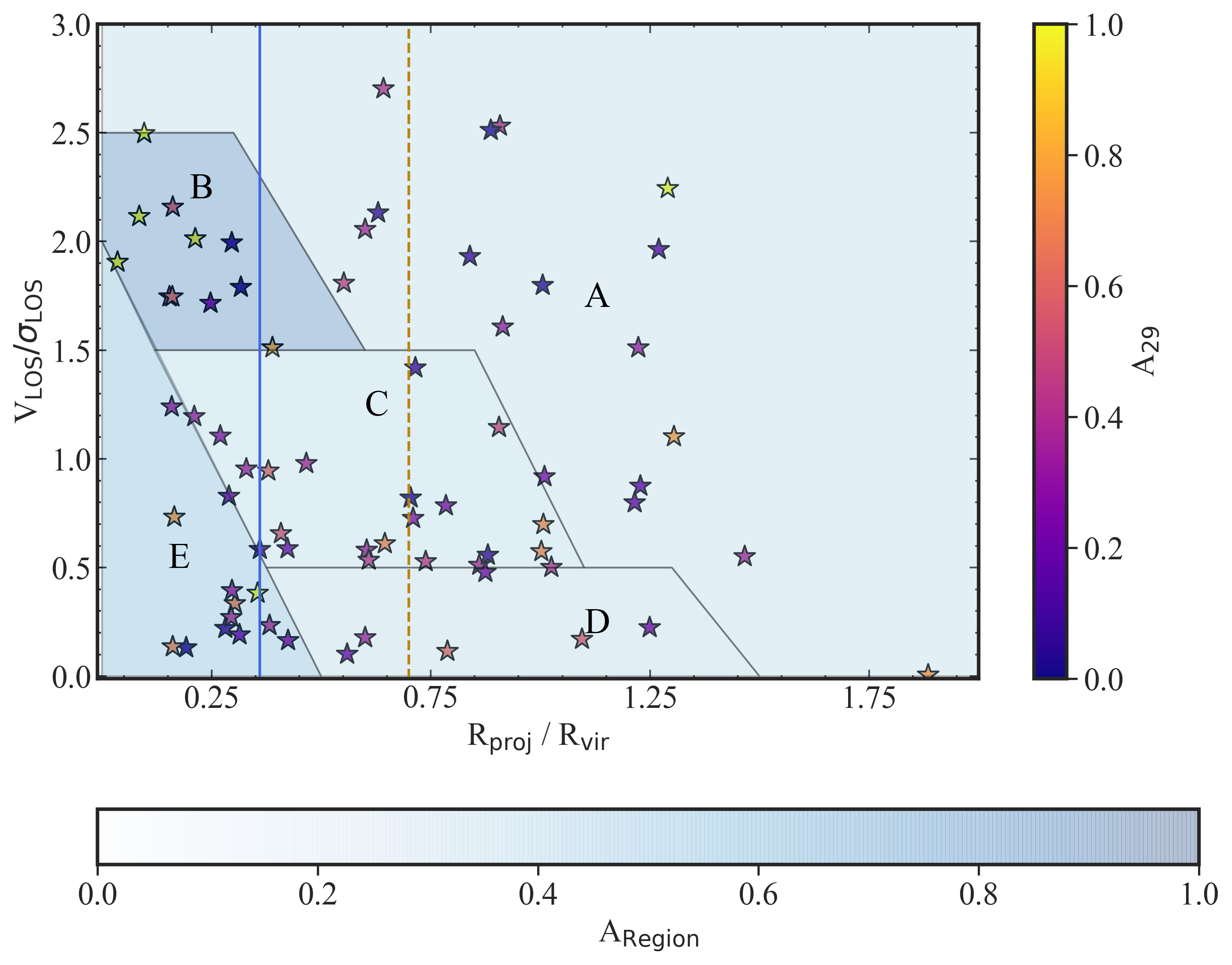}
      \caption{Distribution of Fornax dEs in projected phase-space. The diagram is divided into five regions following the classification by \citet{2017ApJ...843..128R} to indicate the different phase of galaxies infall into the cluster. The colour bar shows the asymmetry value for dwarf galaxies. The region colours represent the mean asymmetry value for galaxies within each respective region.}
         \label{fig:PPS}
   \end{figure*}
   
   Harassment plays a significant role in transforming galaxy morphologies within clusters. Numerical simulations tracking the evolution of harassed galaxies reveal that disturbed spirals are transformed into dwarf ellipticals or dwarf spheroidals (dE/dSph) \citep{1996Natur.379..613M}. This transformation involves high-speed encounters between bright galaxies and Sc-Sd galaxies, resulting in impulsive gravitational shocks that damage the fragile disks of Sc-Sd galaxies \citep{1998ApJ...495..139M}. As the dwarf galaxy falls deeper into the cluster, encounters become more frequent due to higher density of the cluster and the larger number of massive galaxies within it. Consequently, dwarf galaxies in the outskirts of the cluster have more disturbed morphologies, as they have had fewer encounters to disrupt their internal substructure. Furthermore, this process leads to dwarf galaxies being more disturbed as they move towards the cluster centre.

   \xinn{However, it is important to consider the role of dark matter in the dynamics of LSB galaxies. While dark matter does contribute to the mass of dwarf galaxies, its effect is not dominant enough to protect these galaxies from environmental influences. The dark matter fraction in these galaxies is not significantly higher compared to more massive systems. According to kinematic measurements in \citet{2022MNRAS.517.4714E}, the dark matter fraction in dwarf galaxies is at most approximately a factor of five higher than that in more massive galaxies of $10^{10} \, M_{\odot}$, for the faintest dwarfs considered. This indicates that while dwarf galaxies contain a modest increase in dark matter, it only slightly counteracts the influence of environmental effects. Dwarf galaxies, with their lower surface brightness and mass density, are particularly susceptible to environmental mechanisms compared to their more massive counterparts \citep[e.g.][]{2022A&A...660L..11Y,2021MNRAS.507.6045C,2022A&ARv..30....3B}. Additionally, dwarf galaxy harassment further supports this point. Rapid encounters and accreting substructures—harassment—can lead to morphological transformations in dwarf galaxies within cluster environments \citep{1999MNRAS.304..465M}. Therefore, while dark matter does contribute to the mass of dwarf galaxies, its impact is not dominant enough to protect them from environmental effects. Instead, their low mass makes them particularly susceptible to these effects.}
   
   \subsection{The phase-space diagram of Fornax dwarf galaxies}
   The projected phase-space (PPS) diagram, which combines the line-of-sight velocities and projected cluster-centric radii, has become an efficient tool for understanding galaxy evolution inside a cluster and the effect of environmental processes on galaxy properties \citep[e.g.][]{2014ApJ...796...65M,2015MNRAS.448.1715J,2017ApJ...843..128R,2020A&A...635A..36G}. Recently, studies have demonstrated the relationship between a galaxy's phase-space location and its star formation activity \citep{2011MNRAS.416.2882M, 2014MNRAS.438.2186H,2019MNRAS.484.1702P,2019ApJ...876..145S,2020ApJS..247...45R}, infalling history \citep{2013MNRAS.431.2307O,2017ApJ...843..128R}, and the effect of the ram-pressure stripping \citep{2014MNRAS.438.2186H,2015MNRAS.448.1715J,2018MNRAS.476.4753J,2019MNRAS.485.1157B}. The phase space coordinates of galaxies can be used to estimate the time since their infall and infer information about their orbital history \citep{2013MNRAS.431.2307O}. 
      \begin{figure}
   \centering
   \includegraphics[width=1\columnwidth]{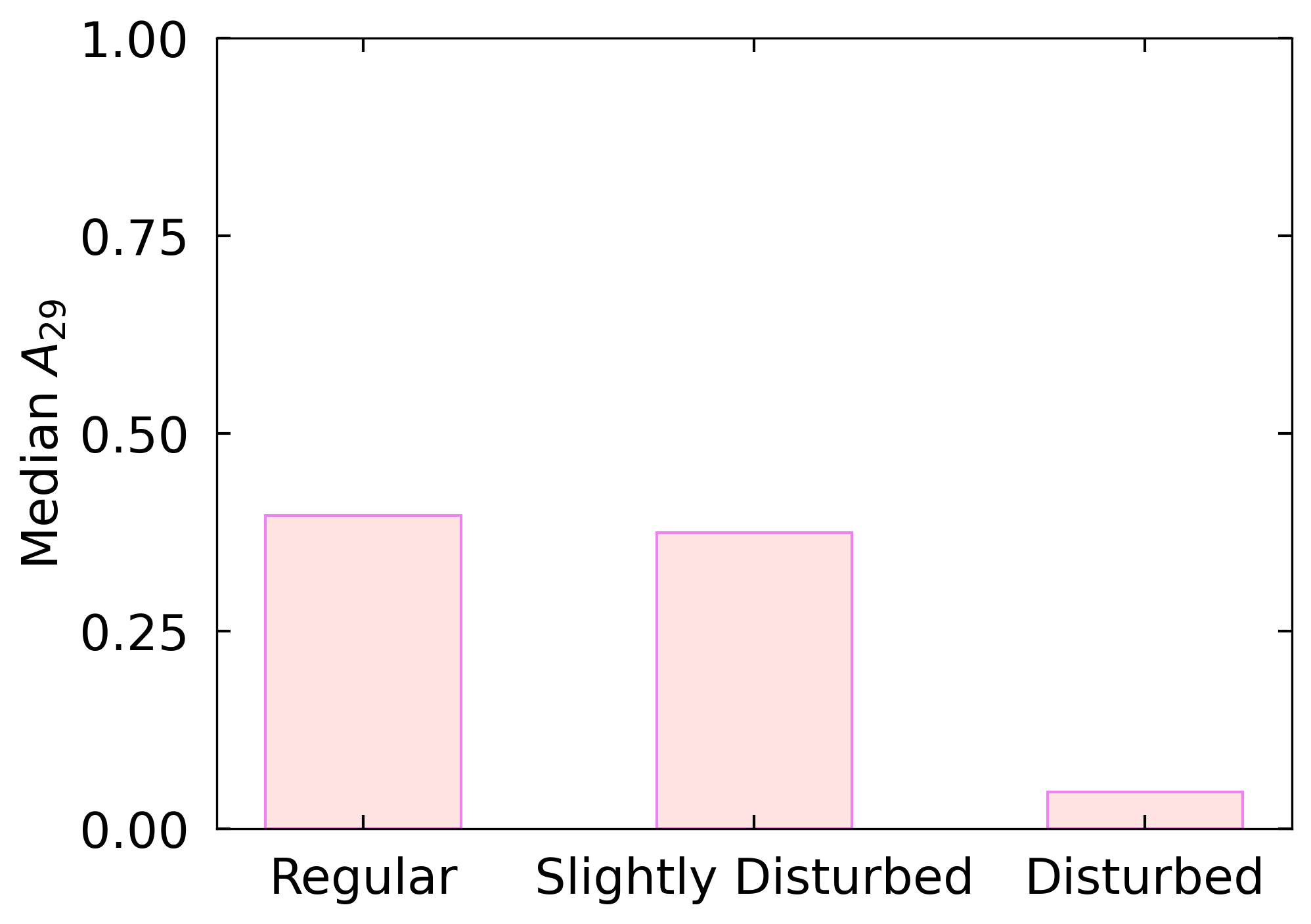}
      \caption{Comparison of median asymmetry values with tidal morphology classes from \citet{2018A&A...620A.165V}. The bars represent the median asymmetry values for galaxies within three primary morphology classes: "regular" (N=390), "possibly disturbed" (N=54), and "disturbed" (N=3), where N is the number of galaxies in each class.}
         \label{fig:A_morph}
   \end{figure}
   \begin{figure}
   \centering
   \includegraphics[width=\columnwidth]{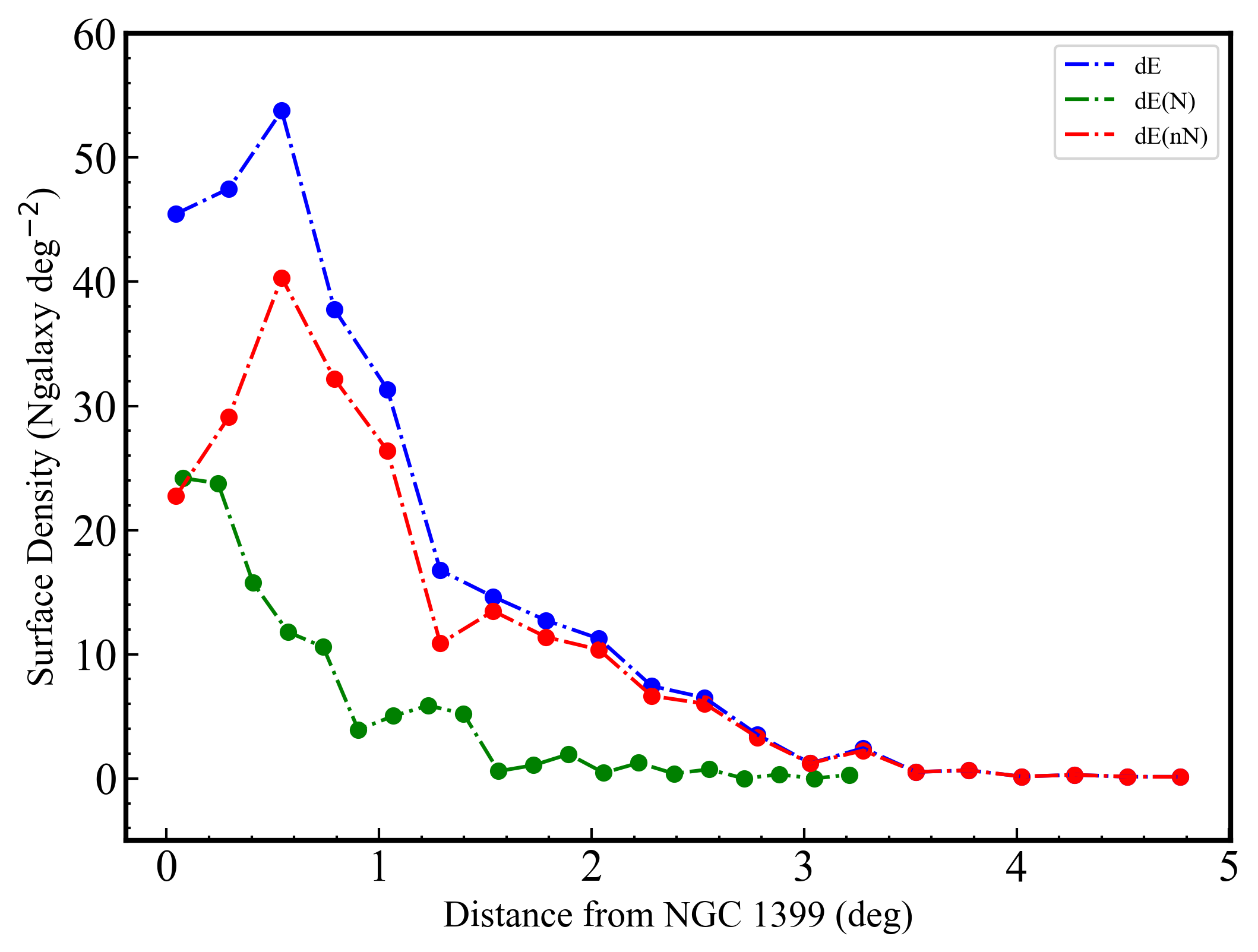}
      \caption{Surface number density of galaxies as a function of distance from the cluster centre in the inner region of the Fornax cluster. The blue dashed line corresponds to early type dwarf galaxies, and the green and red dashed lines represent dE(N)s and dE(nN)s, respectively.}
     \label{fig:fornax_sfd}
   \end{figure}
      \begin{figure}[!htbp]
   \centering
   \includegraphics[width=\columnwidth]{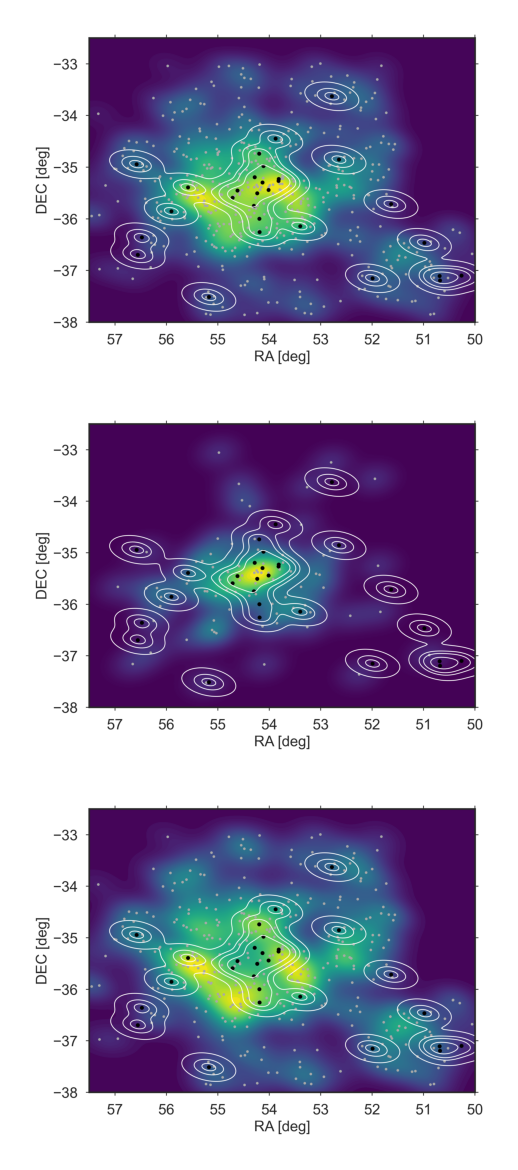}
      \caption{Distribution of dwarf elliptical galaxies in the Fornax cluster shown as surface number density maps. The figure shows the distributions of all dEs (top panel), nucleated dEs (middle panel), and non-nucleated dEs (bottom panel). White contours indicate the distribution of massive galaxies, while black dots represent their specific positions. \xxin{The white contours represent isodensity levels derived from the KDE for massive galaxies, while the black dots indicate their specific positions.} Grey dots mark the positions of dwarf galaxies.}
         \label{fig:fornax_2d}
   \end{figure}
   
   In this subsection, we build the \petra{PPS} diagram of \petra{the} Fornax dwarf galaxies \petra{in our selection} to better understand the assembly history of the galaxy cluster (Fig.~\ref{fig:PPS}). We use the spectroscopic redshifts of each dwarf from the \citet{2019MNRAS.490.1666M} catalogue and velocity measurements from Eftekhari et al. (2022), and crossmatch them with our own catalogue. The projected distance is normalised by the virial radius of 0.7 Mpc, while the velocity is normalised by the cluster velocity dispersion of 303 km/s \citep{2019MNRAS.490.1666M}. Additionally, the cluster galaxies are colour-coded based on their asymmetry value.
   
   Following the analysis by \citet{2017ApJ...843..128R}, we classify different regions in the \petra{PPS} diagram of the Fornax cluster (Fig.~\ref{fig:PPS}). According to \citet{2017ApJ...843..128R}, galaxies in region A are dominated by interlopers and first infallers. Regions B and C contain about 70\% of the recent infallers. Region D has a higher number of intermediate infallers compared to other regions, while region E is dominated by ancient infallers that have fallen into the cluster a long time ago. Region B, dominated by objects near pericentre, shows a higher level of asymmetry. This suggests a potential triggering of asymmetry near the first pericentre as satellites undergo significant disruption. Region D, dominated by backsplash galaxies, does not exhibit an enhancement in asymmetry, which suggests that asymmetries generated at pericentre may not last long. Region E, dominated by dwarfs near the cluster centre, also shows a slightly high degree of asymmetry.
%----------------------------------------------------------------- 
\subsection{The inner regions of the Fornax Cluster}
We have provided a potential explanation for the high asymmetry of galaxies in the outer region of the Fornax cluster in Section 5.1. In this subsection, we aim to discuss the asymmetry in galaxies located in the inner region of the Fornax cluster.

Asencio et al. (2022) study the tidal disruption in dwarfs of the FDS sample studied here. They model the fraction of disturbed dEs in the Fornax cluster, based on a visual study by Venhola et al. (2022), which concludes that the fraction of disturbed dEs is increasing towards the centre of the cluster.  Asencio et al. conclude that the large number of disturbed galaxies cannot be explained by tidal effects within the $\Lambda$CDM theory, as based on the work of Penarrubia et al. (2009) and van den Bosch et al. (2018) and therefore argue that an alternative gravitational models, like MOND, has to be invoked. 

A way out, however, would be if fewer dwarfs were disturbed, or if they had not been detected. This is one of the reasons we have done the current study in a quantitative way. \xxin{We find indeed that galaxies in the very central region are significantly more asymmetric than galaxies further out. This is, however, only true in the very centre, up to 0.12 Mpc (0.33 degrees). Beyond that, we do not detect any significant enhancement in asymmetry. This limiting radius of 0.12 Mpc is consistent with the visual classification \xinn{of \citet[Fig. 3]{2022MNRAS.515.2981A}, based on \cite{2022A&A...662A..43V}}. Thus, our quantitative results
are in good agreement with the results reported in \citet{2022MNRAS.515.2981A}, \xinn{based on
visual classification}. We propose that tidal effects are quite significant in the
very inner regions of the Fornax Cluster, leading to asymmetrical distortions in the outer light profiles of dwarf galaxies.}

In Appendix \xinn{D} we show that the biases associated with not detecting objects in the very center are very limited.

We compare our results with the tidal morphology classified by \citet{2018A&A...620A.165V}, as shown in Fig.~\ref{fig:A_morph}. The x-axis shows three primary classes: "regular," "possibly disturbed," and "disturbed," while the y-axis represents the median asymmetry values for galaxies in each of these classes. We find no correlation between these classes and the asymmetry values. This discrepancy may arise from the visual inspection methods used by \citet{2018A&A...620A.165V}.

There are also other indicators of tidal effects.
   In the densest region of the cluster, where several signs of gravitational interactions have been detected, the evolution of galaxies appears to have been significantly influenced by their environment, including the effect of tidal stripping. A faint bridge-like stellar stream was detected within the intracluster region between NGC 1399 and NGC 1387, indicating an ongoing interaction between these galaxies \citep{2016ApJ...820...42I}. Furthermore, \citet{2017ApJ...851...75I} have identified a previously undiscovered intra-cluster light (ICL) region in the core of the Fornax cluster. This light was shown to be the counterpart of previously detected over-densities of blue globular clusters \citep{2006A&A...451..789B,2016ApJ...819L..31D}, which are indicators of a tidal stream. These findings support the idea that the ICL in the core of the Fornax cluster is formed through the process of tidal stripping, where material, including stars and globular clusters, is pulled away from the outskirts of galaxies during close interactions with the central cD galaxy. \citet{2019A&A...623A...1I} study bright early-type galaxies around the core region of the Fornax cluster, and find that gravitational interactions between galaxies happen within this region, resulting in asymmetries in the structure of the galaxy outskirts. Furthermore, \citet{2016ApJ...819L..31D} discover a complex extended density enhancement in the Globular Clusters (GCs) within the central region of the Fornax cluster, indicating a history of galaxy-galaxy interactions in the core and the stripping of GCs from the halos of core galaxies by the gravitational potential. Also, previous studies suggest that some ultra-compact dwarf (UCD) galaxies in the Fornax cluster core likely originate from the stripping of dwarf galaxies, and they are considered as remnant nuclei of disrupted dwarf galaxies, providing an indicator for galaxy disruption processes \citep{2003Natur.423..519D,2016A&A...586A.102V,2016MNRAS.459.4450W,2021MNRAS.504.3580S}. 
   
   Another piece of information is the radial distribution of dEs in the cluster, as given in Fig.~\ref{fig:fornax_sfd}, constructed using the complete, magnitude limited sample used in this paper. 
   %The exponential fit provides an estimate of the expected original number of dwarf galaxies in the inner region of the cluster, if the exponential fit, which is excellent in the outer parts, would continue inwards. The discrepancy between the expected and observed numbers of galaxies can be calculated by integrating under the exponential curve, giving a predicted value of 36.17 galaxy counts, while the observed data only accounts for 12.02, 
   \xx{This distribution indicates a loss of dwarf galaxies in the inner region of the cluster, suggesting that tidal effects may be playing a role.} Additionally, as shown in the figure, \xx{there has probably been a reduction in the number of non-nucleated galaxies near the cluster centre. }The strong correlation of nucleation fraction with environment (Venhola et al. 2019) is evidence for tidal effects being responsible.

   \xxin{Nucleated dwarf galaxies, which have distinctively bright and compact central nuclei \citep[e.g.][]{2006ApJS..165...57C}, have mainly been studied in the environment of dense clusters \citep{2014MNRAS.445.2385D,2018ApJ...860....4O} and less populated groups \citep{2009MNRAS.396.1075G,2020A&A...634A..53F}. \citet{2021A&A...650A.137F} show that nuclear star clusters (NSCs) form through two main mechanisms: in massive dwarfs, they are mostly the central population that formed from the last burst of star formation, which efficiently builds up the mass of the most massive NSCs. In contrast, in fainter dwarfs, NSCs form from the merging of globular clusters that fall into the centre. It has been shown that nucleated dwarf galaxies are more prevalent near the centres of clusters, such as the Fornax cluster \citep{1989ApJ...346L..53F,2019A&A...625A.143V}.} To further investigate the distribution of nucleated and non-nucleated dEs within the Fornax cluster, we conduct a kernel density estimate (KDE) analysis, as shown in Fig.~\ref{fig:fornax_2d}. The figure shows the distributions of all dEs, nucleated dEs, and non-nucleated dEs from top to bottom. An observable discrepancy is evident between the distributions of nucleated and non-nucleated dEs. Using the positions of massive galaxies from \citet{2021A&A...647A.100S}, we find that nucleated dEs tend to align more closely with the distribution of massive galaxies compared to non-nucleated dEs. The lack of non-nucleated dEs in the central region of the Fornax cluster could potentially be attributed to tidal forces near the cluster centre, which may destroy these galaxies. This observation is consistent with the phase-space diagrams presented in \citet{2019MNRAS.484.1702P}, which also show a reduction in the number of objects near the very center.
   
   %In addition to our analysis of the Fornax cluster, we have extended our investigation to the Virgo cluster using data from the catalogue by \citet{2020ApJ...890..128F}. Fig.~\ref{fig:virgo_sfd} shows the surface number density of galaxies as a function of distance from the cluster centre in the core of the Virgo cluster. Notably, the surface density appears to show a different pattern compared to that observed in the Fornax cluster. Fig.~\ref{fig:virgo_2d} presents a kernel density map of the Virgo cluster members, showing variations in the distribution of galaxies within the cluster central region. 

      %----------------------------------------------------------------- 
   \section{Summary and Conclusions}
   \label{sec:conclusion}

   With FDS data, we investigate the detailed morphological characteristics of dwarf galaxies in the Fornax cluster \petra{by} focusing on the effect of the surrounding environment on their morphology. We quantify their morphologies by calculating the asymmetry (A) and smoothness (S) parameters of 556 dwarf galaxies and analyse their relationship with the distance from the cluster centre. \petra{Our conclusions are as follows:}

   \begin{itemize}
       \item %We find a notable increase in asymmetry among dwarf galaxies at larger distances from the center of the Fornax cluster. 
       %Our findings reveal that dwarf galaxies located in the outer regions of the Fornax cluster have higher values of asymmetry compared to other dwarfs, indicating that these galaxies are, on average, more morphologically disturbed. This result suggests that the surrounding environment plays a significant role in influencing the morphology of dwarf galaxies in the galaxy cluster. 
       \xx{Our findings reveal that dwarf galaxies located in the outer regions of the Fornax cluster have statistically significant higher values of asymmetry compared to other dwarfs, and tend to be bluer. Galaxy colour and asymmetry are also correlated. Therefore, we propose that the outer dwarfs are less quenched by the environment and their bright star forming regions enhance their asymmetry values.}
              
       \item \xxin{We find that the asymmetries in the very inner regions are significantly enhanced, }\xx{probably} due to tidal perturbations. This is supported by a \xx{reduction} of dwarfs in the central regions of the Fornax cluster. Additionally, our observations indicate that in the central region of the cluster, \xx{it appears that some non-nucleated galaxies may have disappeared}, further suggesting the environmental effect on the morphological transformation of these objects.
   
        \item We show that \xxin{fainter} galaxies are more susceptible to environmental effects, such as harassment and tidal stripping, compared to \xxin{brighter} galaxies.
   
        \item Our results \petra{demonstrate} a notable relationship between asymmetry and magnitude. Faint galaxies have higher asymmetry values, indicating that they are more susceptible to the effects of their surrounding environment. On the other hand, a hint of high asymmetry observed in bright galaxies suggests possible influencing factors such as dust.

    \end{itemize}
    In conclusion, our study shows the significant effect of the surrounding environment on the morphology of dwarf galaxies in the Fornax cluster. In future work, we will further analyse the morphologies of more dwarf galaxies within galaxy clusters (e.g. Perseus Cluster) and study the role of the environment\xxin{, using the Euclid Early Release Observations (EROs) data of the Perseus galaxy cluster \citep{2024arXiv240513496C}. We will also refer to the findings of \citet{2024arXiv240513502M}, which identify and characterize the dwarf galaxy population in the Perseus cluster, providing a comprehensive catalogue and insights into their morphologies, nuclei, and globular cluster systems.}

\begin{acknowledgements}
      ......
\end{acknowledgements}

%\end{document}

\begin{appendix} %First appendix

\onecolumn
\section{Radial Distribution of Asymmetry in Fornax Dwarfs}
\begin{figure*}[h!]
   \centering
   \includegraphics[width=\columnwidth]{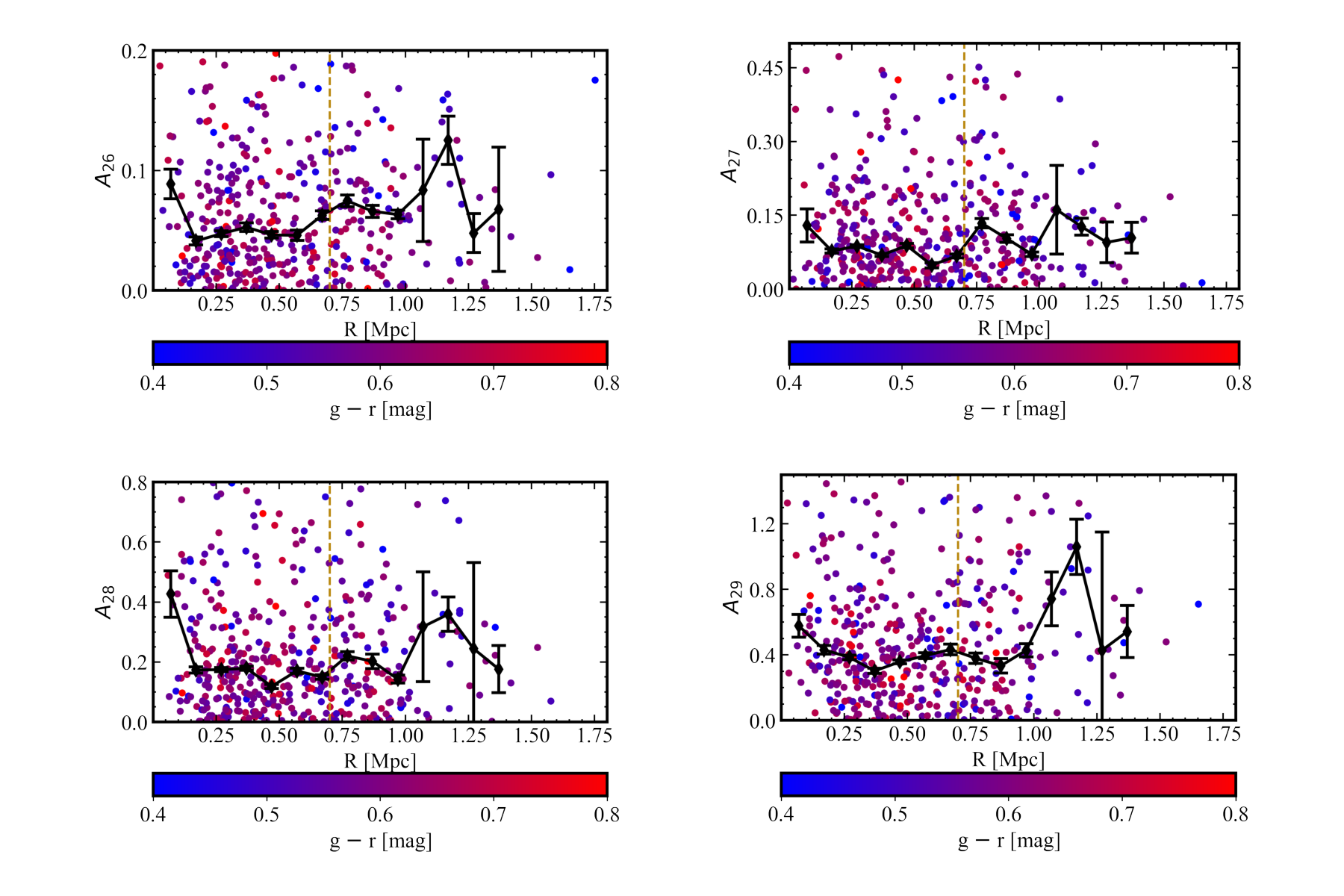}
      \caption{Plots of A versus R (projected distance from the cluster centre). Each panel in the figure corresponds to a specific magnitude range in the $r$ band, as follows:
   Panel (a): Magnitude range: 25.5 - 26.5,
   Panel (b): Magnitude range: 26.5 - 27.5,
   Panel (c): Magnitude range: 27.5 - 28.5,
   Panel (d): Magnitude range: 28.5 - 29.5.}
         \label{fig:A_r_appx}
   \end{figure*}

In Fig.~\ref{fig:A_r_appx}, we present the asymmetry (A26, A27, A28, and A29) as function of the projected distance from the centre of the Fornax cluster. \\
%In Fig.~\ref{fig:A_r_checked_appx}, we present the Plots of A versus R with visual filtering applied.\\
%In Fig.~\ref{fig:A_r_sf_appx}, we present the Plots of A versus R for HSB and LSB galaxies.\\

%\section{Radial Distribution of Smoothness in Fornax Dwarfs} %Second appendix
%In Fig.~\ref{fig:S_r_appx}, we present the Smoothness (S26, S27, S28, and S29) as function of the projected distance from the center of the Fornax cluster. \\
%\begin{figure*}
%\includegraphics[width=\textwidth]{images_0/S_vs_r.png}
%\caption{Plots of S versus R (projected distance from the cluster centre).}
%\label{fig:S_r_appx}
%\end{figure*}
%\section{Four example dwarf galaxies in the cluster central regions}
%Fig.~\ref{fig:image_center_gal} shows several images of dwarfs in the central regions of the Fornax cluster, accompanied by a corresponding table showing the asymmetry, smoothness and colour value for each galaxy. \\
%\begin{figure*}
%\centering
%\includegraphics[width=\textwidth]{images_1/center_gal.png}
%\caption{Four example dwarf galaxies in the central regions of the Fornax cluster with high asymmetry. Additionally, the asymmetries of each galaxy are shown.}
%\label{fig:image_center_gal}
%\end{figure*}

\section{Asymmetry-Magnitude Relationship in Fornax Dwarfs} %Second appendix
\begin{figure*}
   \centering
   \includegraphics[width=\columnwidth]{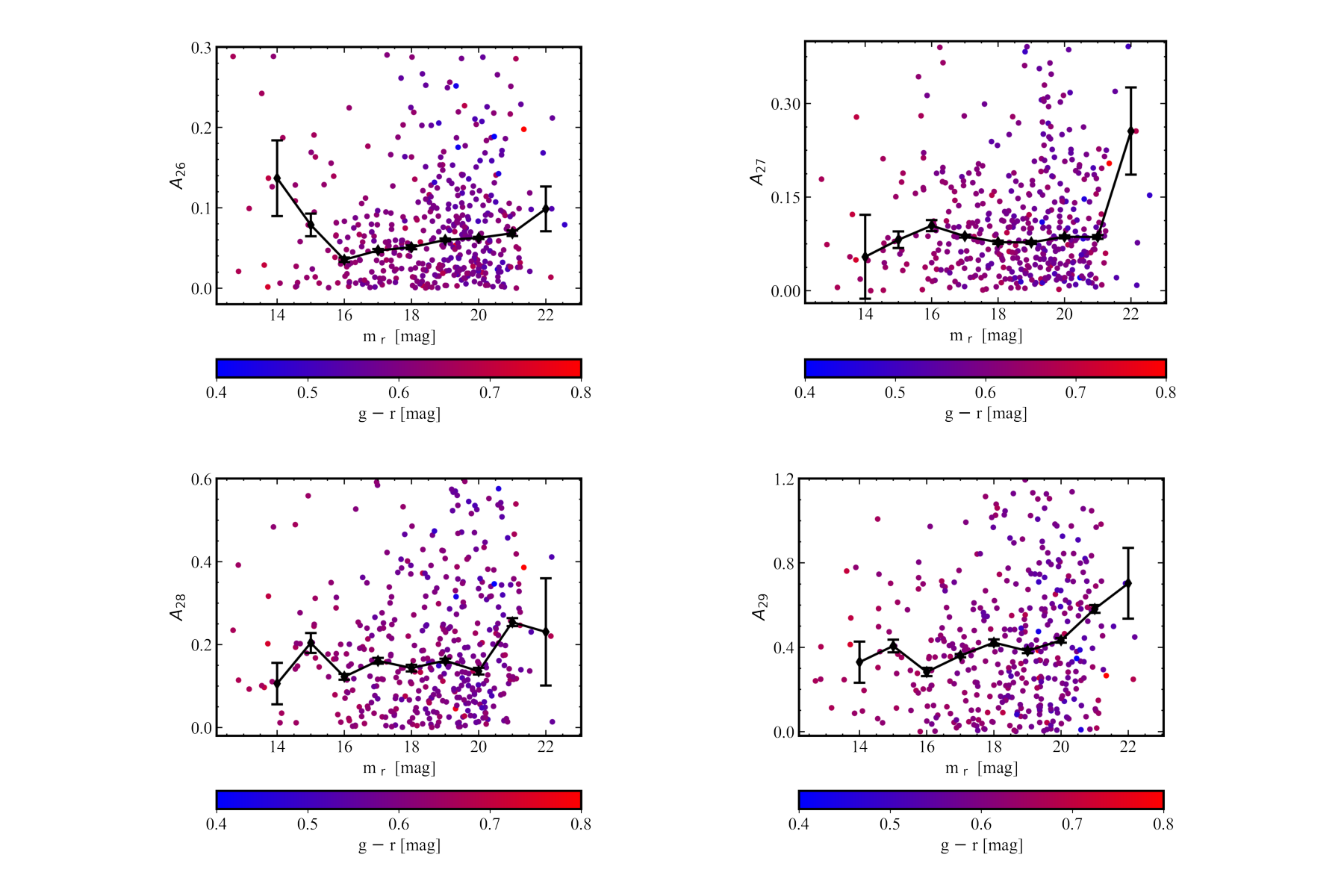}
      \caption{Asymmetry-Magnitude Relationship in Fornax Dwarfs.}
         \label{fig:A_mag_appx}
   \end{figure*}

Figure~\ref{fig:A_mag_appx} presents the relationship between asymmetry and magnitude in the r band for Fornax dwarf galaxies, with four panels displaying A26, A27, A28, and A29. \\
%Fig.~\ref{fig:A_mag_checked_appx} presents the plots of A versus magnitude with visual filtering applied.\\

\section{Smoothness vs. Distance in Fornax Dwarfs}
\begin{figure*}[h!]
   \centering
   \includegraphics[width=\columnwidth]{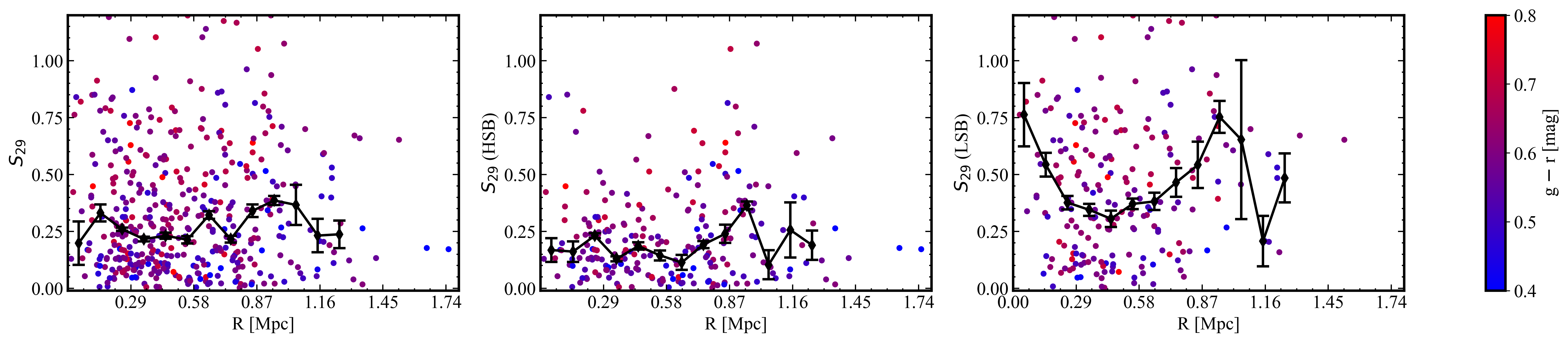}
      \caption{Same as Fig.~\ref{fig:A_r_appx} for the $S$ -- $R$ relation. The three panels represent all galaxies (left), HSB galaxies (middle), and LSB galaxies (right).}
         \label{fig:S_r}
   \end{figure*}

Figure~\ref{fig:S_r} shows the relationship between smoothness and the distance from the cluster centre. %The results indicate that dwarf galaxies located in the outer region, exhibit higher values of $S$ compared to other dwarf galaxies. Additionally, 
We observe an overall higher value of smoothness in LSB galaxies, which suggests they are more easily disturbed. Specifically, 44\% of LSB galaxies exhibit smoothness values higher than 0.5, compared to only 12\% of HSB galaxies. 
  
\section{Detection Ratio vs. Stellar Mass and Distance} 
\begin{figure*}
   \centering
   \includegraphics[width=0.7\columnwidth]{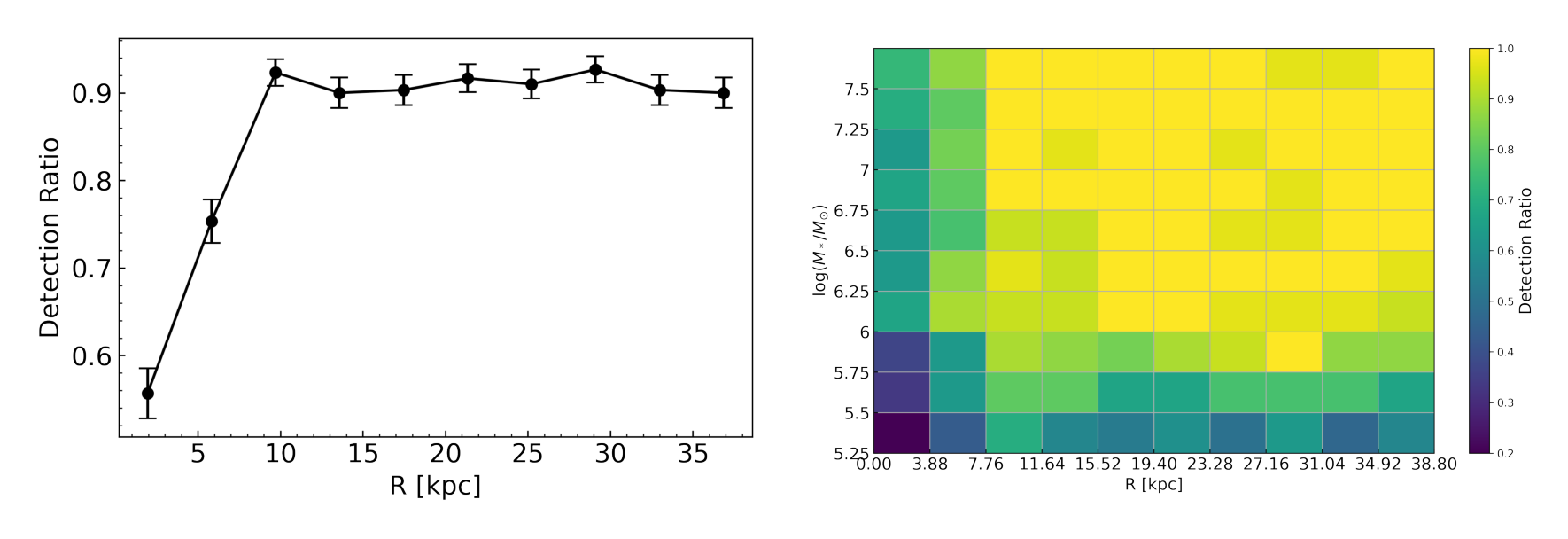}
      \caption{Detection ratio of Fornax dwarf galaxies. The left panel shows the detection ratio as a function of distance from the cluster centre, while the right panel shows the detection ratio in relation to both stellar mass and distance from the cluster centre.}
         \label{fig:det_ratio}
   \end{figure*}

We have done some simulations to find out up to which clustercentric radius the detection of dwarfs is affected by the presence of NGC 1399. For this we took the $r$-band FDS data, injected galaxies of a range of magnitudes, following the scaling relations of \citet{2019A&A...625A.143V} and with a range of ellipticities. Fig.~\ref{fig:det_ratio} shows the detection ratio for dwarf galaxies in the Fornax Cluster. The left panel displays the detection ratio as a function of distance from the cluster centre, while the right panel shows how detection ratio varies with both stellar mass and distance from the cluster centre. We find that there is only a bias in the inner 10 kpc, implying that our results in the paper are not affected.\\

\end{appendix}
\end{document}